\documentclass{article}
\usepackage{geometry,amsmath,amssymb,graphicx}
\usepackage[american]{babel}
\newcommand{\nobracket}{}
\newcommand{\nocomma}{}
\newcommand{\nosymbol}{}
\newcommand{\tmop}[1]{\ensuremath{\operatorname{#1}}}
\newcommand{\tmtextbf}[1]{{\bfseries{#1}}}

\begin{document}

\title{$^{171} \tmop{Yb}^+$ System Stability, $5 D_{3 / 2}$ Hyperfine State
Detection Efficiency and F=2 Lifetime}
\author{ \and   \and }
\maketitle

\begin{abstract}
  A data acquisition system is described that is designed to stabilize cooling
  and probe rates to maximize detection sensitivity and minimize possible
  systematic errors due to correlations between drifting experimental
  conditions and varying drive parameters. Experimental parameters that affect
  the $\tmop{Yb}^{171}$ $5 D_{3 / 2}$ hyperfine state preparation and
  detection efficiency are characterized and optimized. A set of wait times
  for optimal sampling of the $D_{3 / 2} ( F = 2)$ lifetime is chosen and used
  to measure that lifetime with high statistical sensitivity. A systematic
  variation in this lifetime seems to be apparent. The source of the variation
  was not identified, but ion number and cooling rate appear to be ruled out.
  A net determination is made of $\tau = 61.8 \tmop{ms} \pm (
  0.6)_{\tmop{stat}} \pm ( 6.4)_{\tmop{sys}}$ which is significantly longer
  than other measurements of the same quantity. An alternate shelving scheme
  is proposed that would provide S-D state discrimination for Yb even isotopes
  as well as improved sensitivity for D state hyperfine discrimination in odd
  isotopes.
\end{abstract}

\section{Introduction}

Trapped ions provide an excellent platform for making a variety of precision
measurements. In particular, there are renewed prospects for a trapped ion
atomic parity violation experiment[\ref{fortson-ionpnc}] using, for example,
light shifts of the $S_{1 / 2}$ and $D_{3 / 2}$ hyperfine states of $^{171}
\tmop{Yb}^+$ generated by driving the S-D transition[\ref{torgerson-ionpnc}].
Measuring these shifts with sufficient sensitivity requires efficiently
preparing and detecting an ion's spin state. State detection using shelving in
$^{171} \tmop{Yb}^+$ is typically done using $D_{3 / 2}$ hyperfine states and
this state's shorter lifetime and relatively small hyperfine splitting
compromise sensitivity. By carefully characterizing pump and probe rates,
experimental parameters such as pump and probe times can be chosen to optimize
the sensitivity of detecting transitions into or out of particular states to
subsequently improve the sensitivity of light shift measurements.

Connecting the measurements from such an experiment to the quantity $Q_W
(^{171} \tmop{Yb})$ of interest for evidence of physics beyond the standard
model of particle physics will require a number of precise atomic structure
parameters including the S-D quadrupole reduced matrix element. This could be
determined from a measurement of the $D_{3 / 2}$ lifetime and reinforced by
theoretical calculations. This lifetime has been measured in $^{174}
\tmop{Yb}^+$ using other methods to be $52.7 \pm 2.4$ms[\ref{yu-lifetime}].
This is not yet sufficiently precise and the uncertainty appears to be
underestimated. This measurement also only considers collisional quenching for
possible systematic errors, and assumes Poisson counting statistics which may
give a systematic shift in the fit lifetime. The $S - D$ matrix element has
not yet been the target of recent precision calculation methods and existing
calculations of the lifetime vary widely, 41ms[\ref{wilson}] and
74ms[\ref{werth}]. A more precise experimental measurement may motivate
further theoretical studies.

The importance of the shelved state lifetime to sensitivity motivates
considering states other than the $D_{3 / 2} (F = 2)$ that is commonly used.
The very long lived $F_{7 / 2}$ state would be an excellent alternative and
initial work to drive such a shelving transition has been done.

\section{\label{h.nhk4p16a16fe}Measurement Cycle}

Schematically a lifetime measurement of a relatively long-lived excited state
in any system is straight-forward: prepare the system in the excited state,
wait some period of time, then probe the system to determine if a decay
transition occurred. Repeat such a trial as necessary to measure a transition
probability, and repeat the entire procedure for a set of different wait times
to find the time dependence of the transition probability and determine the
lifetime. For trapped ions this probe is done by using shelving to determine
if the ion is in some particular state that can be connected to the initial or
final state of the decay transition.

\subsection{Shelving}

A doppler cooled ion scatters photons from the cooling beam at rates of 10's
of MHz, some fraction of which can be detected with a photo-multiplier tube
(PMT) providing a cooling signal typically on the order of a few thousand
counts per second (kcps), $r_c$. The PMT signal will also generally include a
low background rate, $r_b$, giving a net total signal corresponding to $r_T =
r_b + r_c$. While being cooled the ion cycles through all the states involved
in the cooling process.

Shelving consists of driving the ion to some relatively long-lived state that
is not part of this cooling cycle.[\ref{dehmelt-shelving}] When the ion is in
this shelved state the PMT count rate drops to the background rate, $r_b$,
which ideally is easily distinguished from the total count rate. This provides
a means of efficiently determining if a single ion is or is not in the shelved
state in a time fundamentally limited only by counting statistics. For
long-lived states, fairly high cooling rates, and low background rates
practically perfect detection can be done very quickly.

For $^{171} \tmop{Yb}^+$ the $D_{3 / 2} ( F = 2)$ state that is being studied
here can itself be used as the shelved state. For other states or other kinds
of measurements, the transition to the shelved state can be done in such a way
that only an ion initially in some particular state ends up being shelved,
while a different initial state would remain somewhere in the cooling cycle.

\begin{figure}[h]
  \resizebox{0.9\columnwidth}{!}{\includegraphics{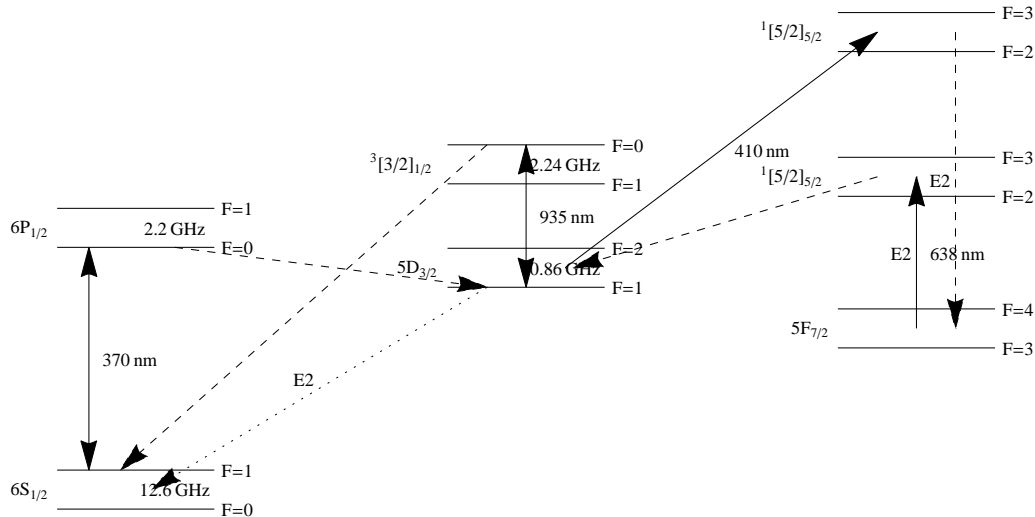}}
  \caption{\label{level-diagram}$^{171} \tmop{Yb}^+$ partial level diagram.
  The states involved in cooling and pump/probe sequences are in the first two
  columns. States involved in the proposed alternate shelving scheme are in
  the third column. Driven transition are shown with solid lines. Transitions
  are E1 unless marked otherwise.}
\end{figure}

To help illustrate the details of using these methods in $^{171} \tmop{Yb}^+$,
a partial energy level diagram is provided in figure \ref{level-diagram}.
$^{171} \tmop{Yb}^+$ is Doppler cooled using the $6 S ( F = 1) \rightarrow 6 P
( F = 0)$ transition driven with a 370nm laser. The P state decays primarily
back to the ground state and has a lifetime of about 8ns so that this
transition when saturated yields a fluorescence rate of \ about 100MHz. The
light collection system detects about $5 \times 10^{- 4}$ of these scattered
photons giving a detected cooling rate of between 2000 and 10,000 counts per
second depending on laser and PMT alignment and the cleanup rates described
presently.

The $P$ state can also decay to the $5 D$ state. The $D$ state has a
relatively long lifetime, $\tau \approx 60 \tmop{ms}$, which would
considerably reduce the cooling and detection rate if permitted only to decay
naturally, so a 935nm laser is used to excite the ion from the $D$ state to a
$^3 [ 3 / 2]_{1 / 2}$ state from which it decays to the ground state. Cooling
can then resume. This 935nm laser is nominally tuned to drive transitions
between the $5 D (F = 1)$ and $^3 [ 3 / 2]_{1 / 2} (F = 0)$ hyperfine levels.
The branching ratio of the P decay to the $6 S$ or $5 D$ state is about 142:1,
so the $D$ state must be cleaned out at a rate greater than about $100
\tmop{MHz} / 142 \lesssim 1 \tmop{MHz}$ to avoid significantly reducing the
cooling rate.

Similarly, though the $6 P (F = 0)$ state can not quickly decay to the $6 S (F
= 0)$ hyperfine level because of angular momentum selection rules, the 370nm
laser may non-resonantly drive the $6 S (F = 1)$ to $6 P (F = 1)$ transition.
The $6 P (F = 1)$ can then decay to the $6 S ( F = 0)$ state also removing it
from the cooling cycle. A 7.4GHz electro-optic modulator (EOM) is used on the
370nm laser to provide (second order) side-bands at 12.6+2.2=14.8GHz to couple
the $6 S (F = 0)$ and $6 P (F = 1)$ states to drive any ion in the former
state back into the cooling cycle. Likewise, the $6 P (F = 0)$ state will not
quickly decay to the $5 D (F = 2)$ state, but an ion in the $6 P (F = 1)$
state, through off-resonant excitation from the directly driven $6 S (F = 1)
\rightarrow 6 P (F = 0)$ transition, or directly from the $6 S (F = 0)$ state
via the EOM induced side-bands, will quickly decay to the $5 D (F = 2)$ state.
Again, an EOM is used to generate $2.24 + 0.86 = 3.1 \tmop{GHz}$ side-bands on
the 935nm laser to couple this state to the $^3 [ 3 / 2]_{1 / 2} (F = 1)$
state and keep the ion in the cooling cycle.

\subsection{Pump and Probe}

The $6 S (F = 0)$ and $5 D (F = 2)$ hyperfine states are relatively dark when
their corresponding EOMs are switched off, making either suitable for use as a
shelved state. Consider the $5 D (F = 2)$ state in particular. With the 935nm
laser's 3.1GHz EOM switched off, this state becomes (nominally) isolated from
the cooling cycle. An ion in this state will yield a count rate equal to the
background rate rather than the much larger cooling rate, providing a probe of
the ion's state. An ion can be driven into this state in the same way though
the rate would be small if relying only on off-resonant couplings, so in
practice a 2.2GHz EOM is also employed on the 370nm laser, coupling the $6 S
(F = 1)$ state to the $6 P (F = 1)$ state from which it can decay quickly to
this $5 D (F = 2)$ shelved state, providing the pump step.

A sequence of these particular beam combinations then simply allows a
measurement of the $D_{3 / 2}$ lifetime. The ion is pumped to the $5 D_{3 / 2}
(F = 2)$ state using the combination of EOM states described, then all the
beams are turned off and the ion is left in the dark for some period of time
during which it may decay to the 6S state. Probe beams are then applied and
the number of counts is recorded.

\subsection{Measurement Sequence}

\begin{figure}[h]
  \begin{tabular}{c}
    \resizebox{0.95\columnwidth}{!}{\includegraphics{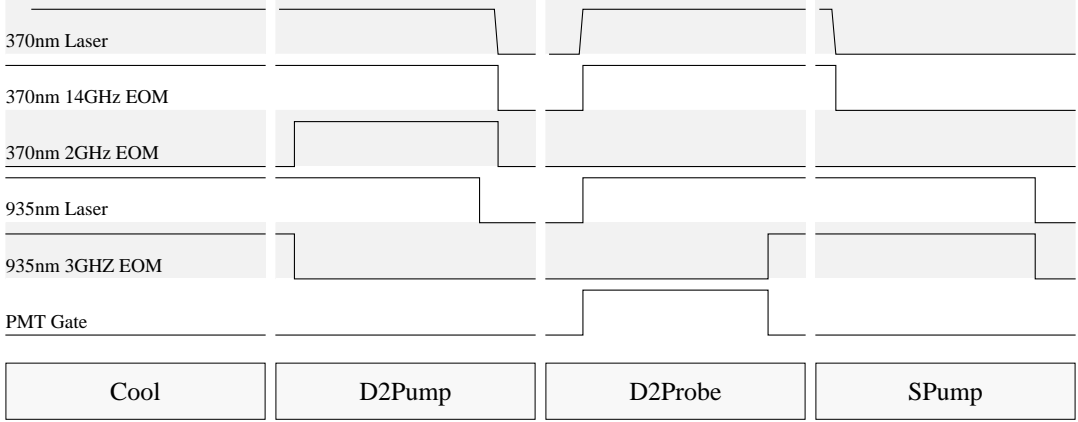}}
  \end{tabular}
  \caption{\label{measurement-units}$^{171} \tmop{Yb}^+$ measurement sequence
  units. These units are combined in blocks, shown in figure
  \ref{measurement-blocks}, that each constitute a single basic measurement.
  In practice the PMT signal is replaced by a pulse at each edge as a PMT is
  triggered rather than gated, and the actual leading and trailing edges of
  the 370nm laser gate are shifted to compensate for the lag of the shutter
  that actually implements this switch. The on and off delays are measured in
  a separate procedure.}
\end{figure}

\begin{figure}[h]
  \resizebox{0.95\columnwidth}{!}{\includegraphics{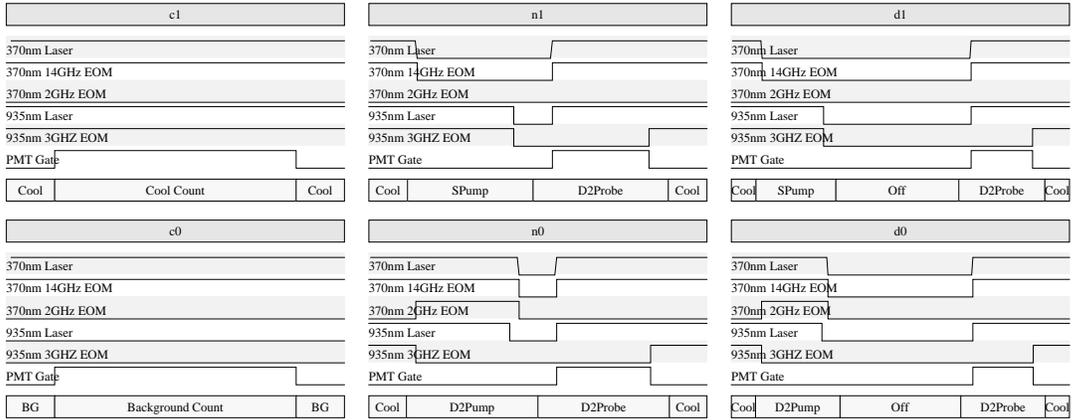}}
  \caption{\label{measurement-blocks}$^{171} \tmop{Yb}^+$ measurement sequence
  blocks formed from units shown in figure \ref{measurement-units}. Trials are
  fixed sequences of these blocks. Parameters such as pump/probe/wait times
  are chosen before each set of trials and a digital waveform is computed
  which is written to a set of digital outputs timed by a hardware clock to
  control the various gates.}
\end{figure}

Figures \ref{measurement-units} and \ref{measurement-blocks} show the basic
building blocks of such a sequence. The pump and probe procedures just
described are referred to as D2Pump and D2Probe, and all beams off during the
wait as Off. Each of these units may be further defined by some parameters,
such as the pump, probe and wait times. Where parameters vary, or are
otherwise of interest they will be included explicitly so that the full
lifetime sequence can be denoted $D 2 \tmop{Pump} ( t_{\tmop{pump}}) /
\tmop{Off} ( t_{\tmop{wait}}) / D 2 \tmop{Probe} ( t_{\tmop{probe}})$. This
entire sequence is in turn denoted simply as $d_0 ( t_{\tmop{wait}})$.

Note that a particular unit is not necessarily a single set of beam states.
D2Pump for example must shut the beams off in a particular way to ensure that
the ion is left in a well defined state. Ideally beams are shut off
instantaneously and simultaneously, in practice each change in beam state
includes delays and transition times that will effectively guarantee that they
are not changed simultaneously. If the 370nm laser is shut off before the
935nm laser, the ion may be driven back to the ground state no matter what the
steady state pumping probability is. Instead the 370nm beam is explicitly shut
off after the 935nm beam in a way that accounts for acousto-optic modulator
(AOM) transition times and shutter lags and all other relevant complications,
and the resulting pump state is thus well defined and stable. Similar
considerations are required for other units and these are reflected in the
timing diagrams for each unit but will not be explicitly discussed here.

A number of other sequences of states are also used in actual experiments to
provide complementary information or to monitor or stabilize various
experimental parameters. For example, SPump is used to drive the ion to some
ground state hyperfine level for the block $d_1 ( t_{\tmop{wait}}) =
\tmop{SPump} / \tmop{Off} ( t_{\tmop{wait}}) / D 2 \tmop{Probe}$ to check for
possible perturbations that result in $6 S \rightarrow 5 D$ excitations even
when all the beams are supposed to be off. And $c_1 = \tmop{CoolCount} (
t_{\tmop{coolcount}})$ is used to stabilize the frequency of the 370nm laser
and provide an independent measure of the cooling rate and its fluctuations.
Others blocks will be discussed as they are used. The counts resulting from a
sequence block are referred to by the name of the block, so that $d_0$ is the
counts resulting from the probe step of a $d_0 ( t_{\tmop{wait}})$ block. A
complete experiment is a multiplexed sequence of these blocks with parameters
and relative frequencies appropriate to maximize the overall sensitivity for
the parameter being measured[\ref{schacht-optimal_sampling}] and minimize the
possible systematic effects of other varying experimental conditions. In most
cases here, a fixed chain of sequences is repeated, such as $c_1 / n_0 / n_1 /
d_0 / d_1 / c_0$, and the freedom to change the relative frequencies of the
blocks is not yet used.

\subsection{State Detection and Sensitivity}

Most sequences end with a $D 2 \tmop{Probe}$ unit which yields a PMT count.
Ideally these counts from a single probe could be used to determine with near
certainty if a transition was made. After a number of trials the probability,
$s$, that the ion is in the shelved state at the beginning of the probe can be
measured, and the lifetime $\tau_D$ determined from
\begin{eqnarray*}
  s ( t_{\tmop{wait}}) & = & s_{\tmop{pump}} e^{- t_{\tmop{wait}} / \tau_D}
\end{eqnarray*}
with $s_{\tmop{pump}}$ the probability for the ion to be in the shelved state
after pumping and $t_{\tmop{wait}}$ the wait time between the pump and probe
stages. The uncertainty in the measurement of $s$ would be given by binomial
statistics.

For the case of shelving using the $5 D_{3 / 2} (F = 2)$ state this is less
straight-forward for reasons that will be considered in detail below. Instead,
the average number of photons $n$ collected during the probe as a function of
wait time is measured. $n$ can be linearly related to $s$ through two
parameters $\bar{n}_b$ and $\bar{n}_c \nosymbol$. With $\bar{n}_T \equiv
\bar{n}_b + \bar{n}_c$
\begin{eqnarray*}
  n & = & s \bar{n}_b + ( 1 - s) \bar{n}_T = \bar{n}_T - \bar{n}_c s
\end{eqnarray*}
$\bar{n}_c$ and $\bar{n}_b$ are modified cooling and background count rates
and reduce to the corresponding actual count rates $n_c$ and $n_b$ in the the
limit of probe times short compared to the probe coherence time. The exact
values can be related to various experimental parameters such as these actual
cooling and background rates, probe time, and pump, probe and decay
rates[\ref{schacht-shelving}], but in practice they are more conveniently
directly measured experimentally.

Take $\sigma_s$ to be the uncertainty with which $s$ can be determined in a
single trial, and $\sigma_s^{( N)}$ the uncertainty of $s$ determined from $N$
independent trials so that $\sigma_s^{( N)} = \sigma_s / \sqrt{N}$. This can
be related to $\sigma_n$, the variance of $n$ for a single trial measurement
with simple error propagation
\[ \sigma_s^{( N)} = \frac{\sigma_s}{\sqrt{N}} = \left| \frac{\partial
   n}{\partial s} \right|^{- 1} \frac{\sigma_n}{\sqrt{N}} = \frac{1}{\sqrt{N}}
   \frac{\sigma_n}{\bar{n}_c}^{} \]
For a total observation time $T$, \ and $t_{\tmop{trial}}$ the time needed to
do one trials, the total number of trials would be $N = T / t_{\tmop{trial}}$
giving
\[ \sigma_s^{( N)} = \begin{array}{l}
     \sqrt{\frac{t_{\tmop{trial}}}{T}} \frac{\sigma_n}{\bar{n}_c}
   \end{array} \]
The single trial measurement uncertainty, $\sigma_n,$is given by multi-modal
statistics [\ref{schacht-shelving}]. $\bar{n}_c$ and $\sigma_n$ and
$t_{\tmop{trial}}$ all generally depend on the probe time $t_p$ in a
non-trivial way but $\sigma_s$ can be approximately minimized by choosing $t_p
\approx \tau_{\tmop{probe}} / 2$ where $\tau_{\tmop{probe}}$ is a probe
coherence time that can be determined experimentally.[\ref{schacht-shelving}]

For the particular case of a lifetime measurement the precision with which the
lifetime $\tau_D$ is determined can be estimated from
\begin{eqnarray*}
  \sigma_{\tau} & \approx & \left| \frac{\partial s}{\partial \tau_D}
  \right|^{- 1} \sigma_s \equiv f_{\tau} \tau_D \sigma_s\\
  f_{\tau} & = & \frac{1}{s_{\tmop{pump}}}  \frac{e^{t_{\tmop{wait}} /
  \tau_D}}{t_{\tmop{wait}} / \tau_D} 
\end{eqnarray*}
for a single trial, or with $\sigma_s \rightarrow \sigma_s^{( N)}$ for $N$
trials. The lifetime $\tau_D $ has been included in $\sigma_{\tau}$ explicitly
to exhibit its general scale and make $f_{\tau}$ dimensionless. This result is
strictly correct only when the only parameter to be determined is $\tau$. An
optimal sampling analysis that accounts for the need to also determine the
$\bar{n}_i$ can be used to provide an exact result for
$f_{\tau}$.[\ref{schacht-optimal_sampling}] For present purposes note only
that $\sigma_{\tau}$ is improved as $s_{\tmop{pump}}$ \ increases so that
$s_{\tmop{pump}}$ should be made as large as possible.

\subsection{System Stability and Calibration}

The probe counts are linearly dependent on $s$ so that they can be used to
determine the dynamical quantities of interest. These average result of these
probe counts will evolve between $\bar{n}_T$ and $\bar{n}_b$. It will be seen
in detail that these quantities and others like $s_{\tmop{pump}}$ are
determined by a number of pump and loss rates, all of which depend on the
intensity and tuning of the cooling lasers. Variations in these experimental
conditions can result in extra contributions to the uncertainties of
parameters measured from these quantities which reduces sensitivity or give
possible systematic shifts of those measured parameters. As a result these
rates should be made as stable as possible, and what can't be stabilized must
be monitored either for use in compensating for the remaining fluctuations and
possibly correcting possible resulting systematic errors, or for screening out
data taken during excessively fluctuating conditions.

Laser intensity will depend on laser power and losses during delivery, both of
which are fairly stable; and positioning of the beam on the ion, which is
stable, but not easily controlled. Similarly frequencies of the cooling and
re-pump lasers are stabilized but the absolute tuning is not controlled. This
results in different, arbitrary but stable, values of $\bar{n}_c$ and
$\bar{n}_b$ \ so that probe results for blocks like $d_0 ( t_{\tmop{wait}})$
or $n_0 ( t_{\tmop{pump}})$ can not be compared directly between different
runs when beam position is adjusted or frequencies are re-tuned and re-locked,
but generally once $\bar{n}_i$ has been determined for each run, comparisons
are valid.

\begin{figure}[h]
  a)\resizebox{0.45\columnwidth}{!}{\includegraphics{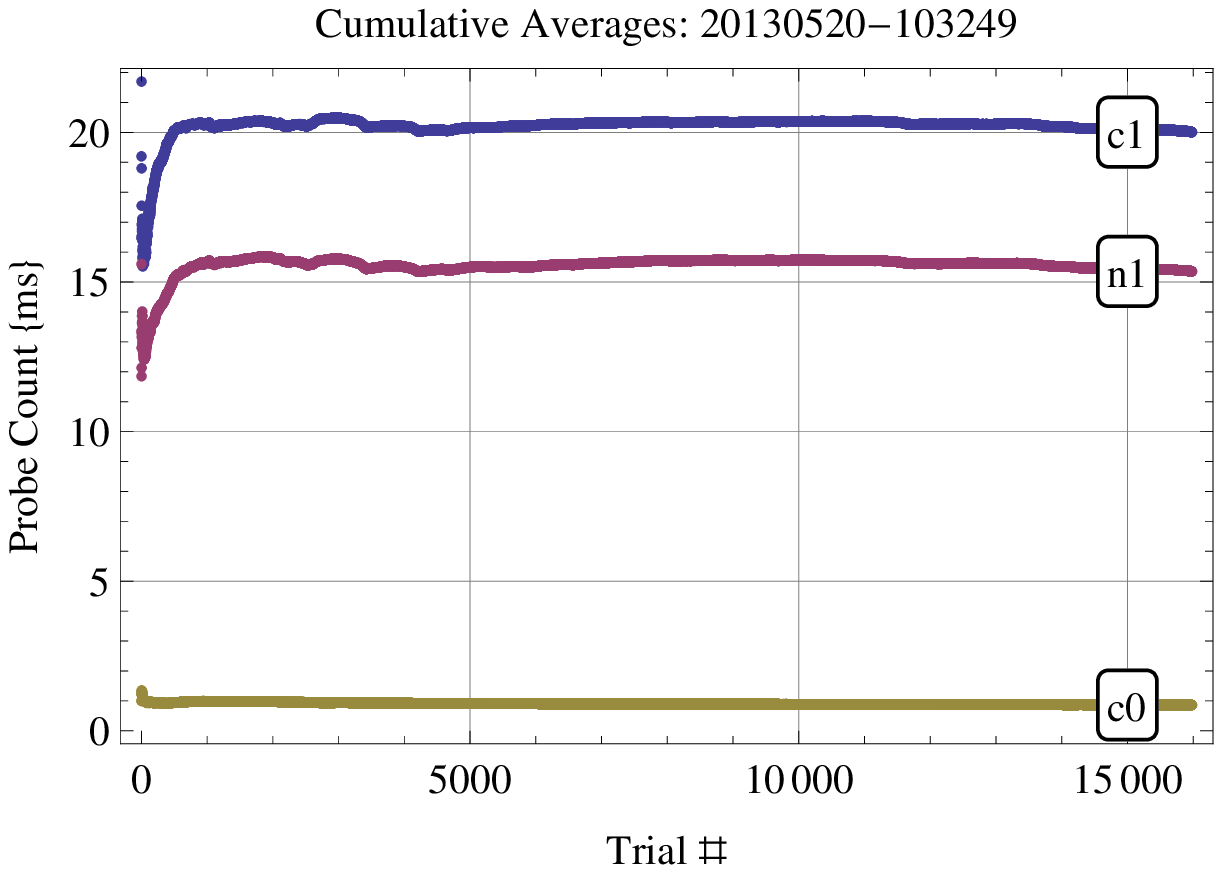}}b)\resizebox{0.45\columnwidth}{!}{\includegraphics{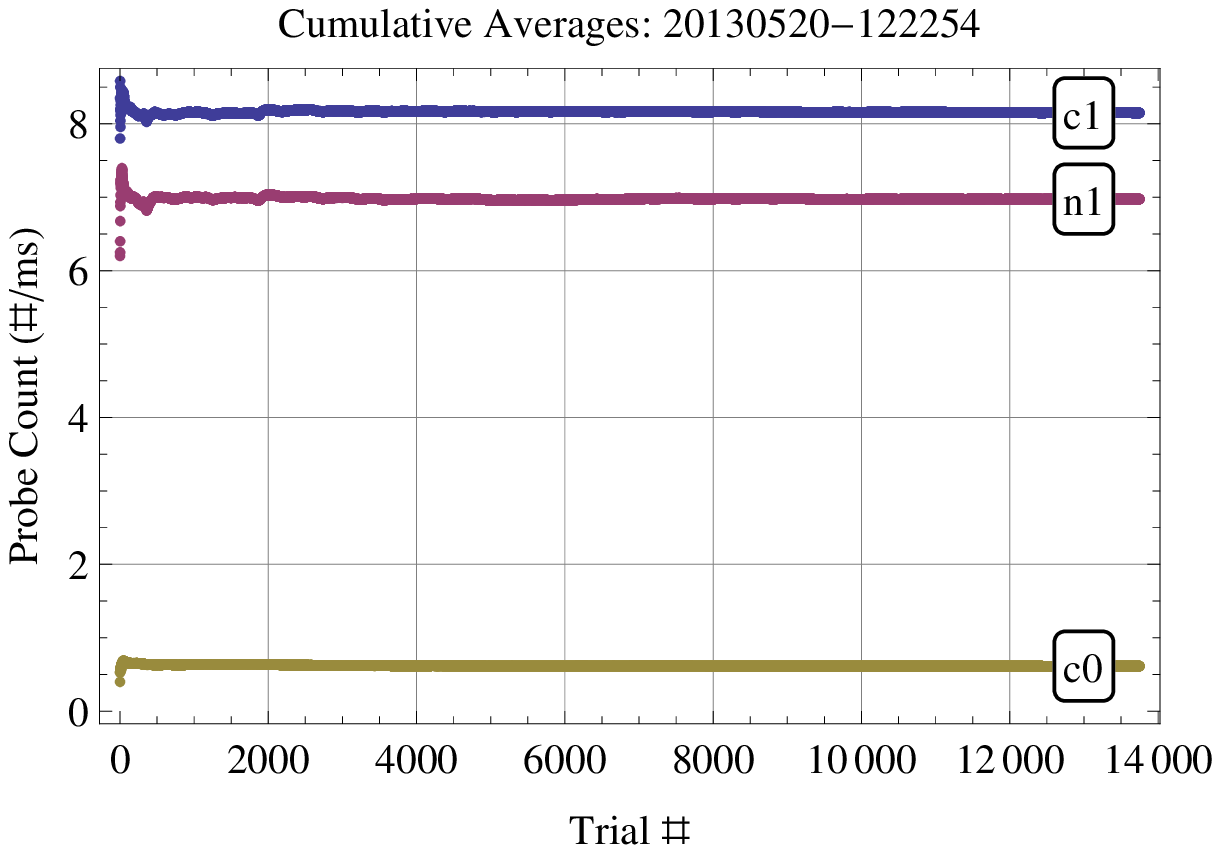}}
  
  c)\resizebox{0.45\columnwidth}{!}{\includegraphics{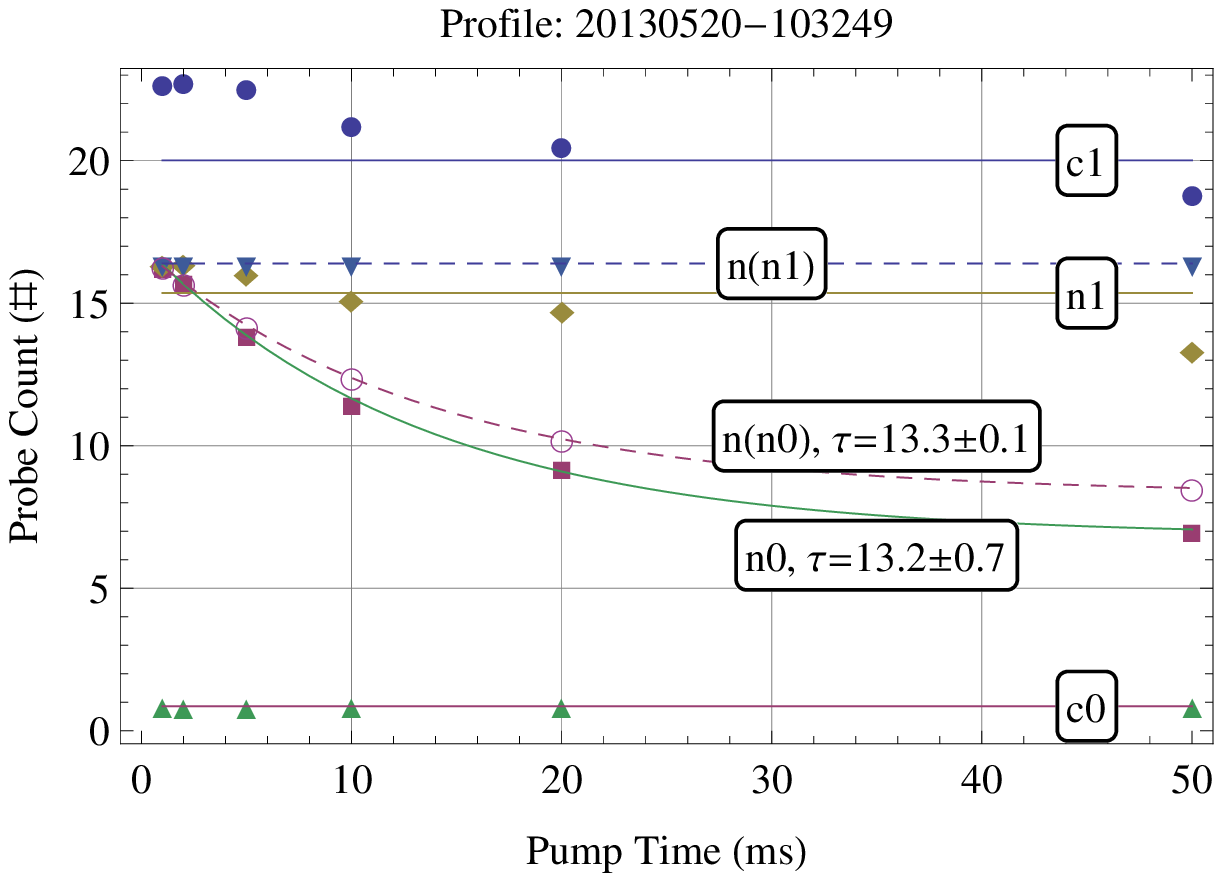}}d)\resizebox{0.45\columnwidth}{!}{\includegraphics{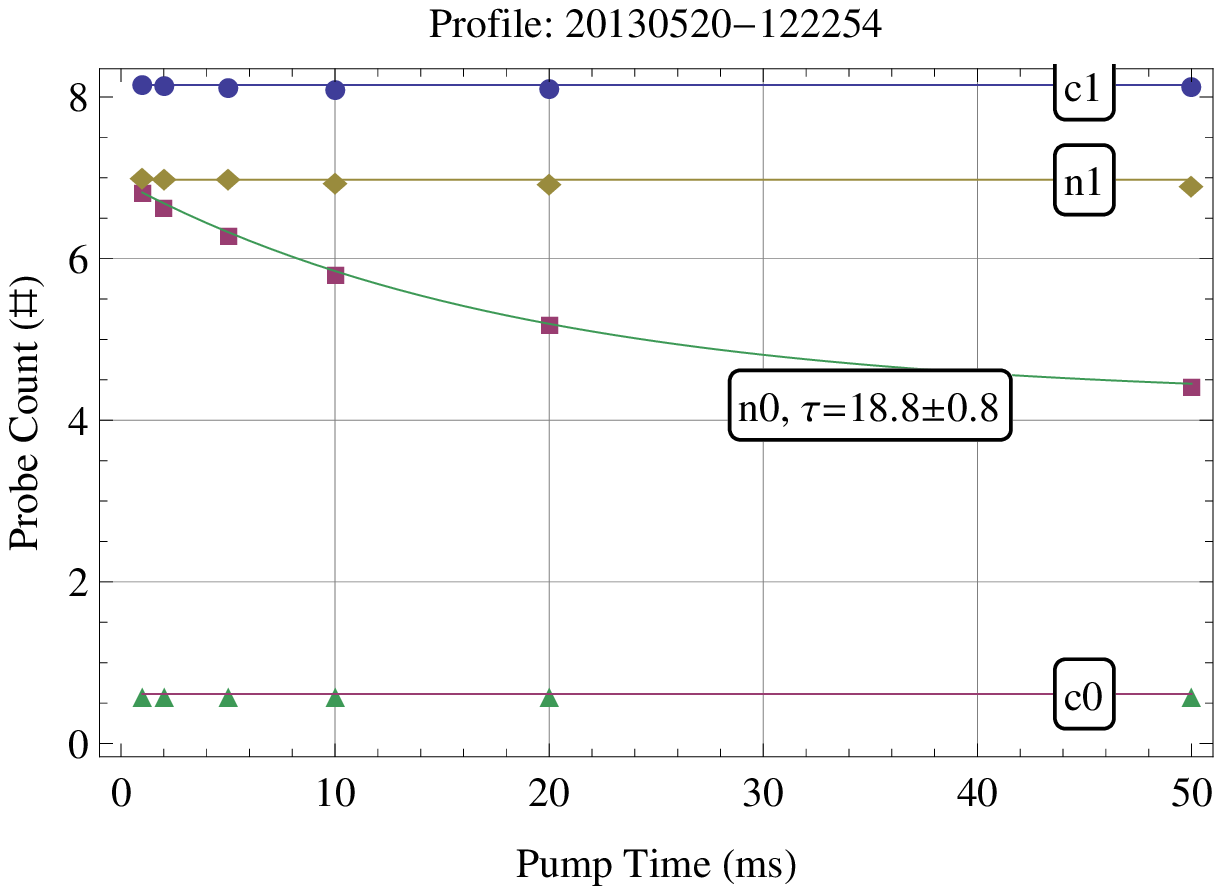}}
  \caption{\label{stability}Examples of data accumulated from the kinds of
  measurements sequences described. This is a probe time measurement. The top
  plots, a) and b) are cumulative averages, as a function of time, of the
  probe counts resulting from the block that is labeled, independent of any
  parameters. This shows the statistical progress of the measurement and
  indications of instabilities. Note that a) shows more variation of $c_1$ and
  $n_1$ over the course of the experiment. The plots in c) and d) are profiles
  of the corresponding probe count as a function of some parameter, in this
  case pump time. This particular example in c) shows a dependence of the
  cooling rate $c_1$ on the pump time, possibly indicating insufficient
  cooling rate. By recording $c_1 ( t_{\tmop{pump}})$ this effect is detected
  and can be used to re-normalize $n_0$ and possibly correct for this effect.
  This is not done for precision measurements. Instead adjustments are made
  until $c_1$ is independent of pump time, as seen in d), the previous data is
  discarded, and the experiment is restarted.}
\end{figure}

For these purposes every measurement sequence includes a number of blocks that
provide information that can be used to determine the $\bar{n}_i$ and assess
or improve their stability. The timing sequences of these additional blocks
are also shown in figure \ref{measurement-sequence}. Among them are $c_1$
giving a direct measure of the cooling rate which is some combination of 370nm
and 935nm rates; $c_0$ giving the background rate; $n_1$ which will be seen to
directly provide $\bar{n}_c$; and $n_0$ giving an upper bound for $\bar{n}_b
\nosymbol$. Particular measurements may also include still other blocks that
provide complementary information about other possible systematic effects, as
will be seen in the lifetime measurement. These values are monitored on strip
chart plots like the example shown in figure \ref{stability}a with either a
cumulative average, as shown, or a shorter timescale boxcar average as
appropriate to facilitate detection of significant variations. The core of the
system used to make these measurements is described in [\ref{torgerson-yb2+}].

Pump and probe rates are most sensitive to the cooling laser frequencies which
are also the most inherently volatile. These are stabilized to the ion itself
using $c_1$. The 370nm laser frequency is updated after every set of trials to
give some fixed average for $c_1$ by re-tuning by an amount proportional to
the difference between $c_1$ and the target value. This can not distinguish
between 370nm or 935nm laser frequency fluctuations, so every 100s or so the
935nm laser is tuned independently outside of the measurement cycle by setting
the frequency to that which maximizes the cooling rate. This stabilization
scheme results in $c_1$ distributions that are almost perfectly Poissonian as
will be seen when considering the probe count distributions (figure
\ref{probe-histogram}a).

This approach yields a very stable overall system that can be used to take
data up to 30 hours unattended. The principle limit to this measurement time
is currently 370nm laser mode quality which seems to deteriorate with time and
temperature and eventually results in insufficient power at the cooling
frequency. When this occurs the system is unable to restore $c_1$ to its
target value by re-tuning the 370nm laser frequency. The data sequence then
stops until the laser diode current and grating piezo voltage are manually
adjusted to restore the cooling rate.

This scheme still allows for certain collective variations of the related
relative rates, such as a 370nm laser power reduction compensated for by
re-tuning, that maintains $c_1$ but yield changes in $\bar{n}_i$, so $n_0$ and
$n_1$ are continuously measured. These are not yet actively stabilized, but
are monitored for sudden or large variations that merit intervention.

Slow drifts of these values can generate systematic errors in certain
measurement. Consider a lifetime or pump/probe rate measurement during which
$\bar{n}_c$ is slowly decreasing due to a cooling laser frequency drift. If
measurements for each particular wait or drive time are all done at once, and
in order of increasing time, probes made during the end of the run would show
a relative reduction having nothing to do with the dynamics of the process
being studied. If this variation is large enough it might be detected by a
poor fit to the expected functional form, but the drift might also combine
with the dynamics to give the same functional form, or just be too small to
identify in this way, and instead lead to a shifted fit parameter. This effect
might be mitigated by normalizing the probe results for each time by the $n_0$
and $n_1$ collected during the same time, but this approach is avoided for
precise measurement due to some lingering uncertainties in the detailed
understanding of the probe process.

More effective is to vary the way the data is collected. By doing smaller
sets of trials for each time, but repeating that sequence many times during
the measurement in a generalized AB pattern, systematic shifts due to linear
drifts will average out. For more general variations, more sophisticated
higher order alternations could be made, e.g. (ABBA)(BAAB), but just as
effective and simpler is to pick a parameter value randomly for each trial. In
this case the measurement times are completely uncorrelated with any long
timescale drifts and all systematic shifts at any order should average out.
Measurements presented here are all done in this way. Parameter values are
chosen randomly from a corresponding predefined set before each set of trials
and a digital waveform that drives all the instruments is generated from these
values. Trial times for a complete sequence of blocks are around 50-200ms and
$\sim 10$ trials are made for each set of parameters so that an actual
measurement consists of a long sequence of 1-2s long sets of trials.

There remains at least one more systematic problem that is occasionally
encountered. Clouds consisting of large numbers of ions experience more driven
motion than small clouds due to the applied radio-frequency trapping field.
During long wait, pump or probe times, when the ions are not being actively
cooled, they may heat up enough that the cooling count rate is reduced due to
increased cloud volume and cooling transition doppler width, and through this
the $\bar{n}_i$ are also reduced or otherwise changed. This gives exactly the
kind of spurious $t_{\tmop{wait}}$ dependence just described. Figure
\ref{stability}a shows just this sort of effect in a pump time measurement.
This can be dealt with in way similar to the case of drifting parameters. The
$n_0 ( t_{\tmop{pump}})$ data can be re-normalized using $\bar{n}_i (
t_{\tmop{pump}})$ and the resulting profile fit. Figure \ref{stability}c shows
the resulting $\tau_{\tmop{pump}}$ for both cases, illustrating the possible
systematic error.

But again, given the unknown, and probably unstable, details of the effects
of the increased heating, in preference to this the system is tweaked until
this variation is eliminated and the $n_0 ( t_{\tmop{pump}})$ are fit
directly. In this case the number of ions trapped was reduced, the 370nm laser
re-tuned to provide a lower $c_1$, and the measurement was repeated. Figure
\ref{stability}d shows that the $t_{\tmop{pump}}$ dependence of $n_0$ has
apparently been eliminated and $c_1$ and $n_1$ in figure \ref{stability}b are
clearly much more stable. The laser re-tuning likely changed pump and probe
rates as well, so the resulting $\tau$ can not be compared directly to the
previous case.

\section{\label{h.n8gs7s9itcvx}Pump Efficiency}

After pumping, the probability that the ion is in the shelved state is
determined by the pump time, $t_{\tmop{pump}}$ and the pump and loss rates,
$\Gamma_{\tmop{pump}}$ and $\Gamma_{\tmop{loss}}$, into and out of the shelved
state. The intermediate states involved are short-lived, and the system can be
described with a first order rate equation
\begin{eqnarray*}
  s_{\tmop{pump}} & = & s_0 e^{- t_{\tmop{pump}} / \tau_{\tmop{pump}}}_{} +
  s_{\infty} ( 1 - e^{- t_{\tmop{pump}} / \tau_{\tmop{pump}}})
\end{eqnarray*}
with $\tau_{\tmop{pump}} = 1 / ( \Gamma_{\tmop{loss}} + \Gamma_{\tmop{pump}})$
and $s_{\infty} = \Gamma_{\tmop{pump}} / ( \Gamma_{\tmop{pump}} +
\Gamma_{\tmop{loss}})$. The quantity $s_0$ is the probability to be in the
shelved state when pumping begins. Pumping is usually done immediately after
cooling. During cooling, transitions into the $5 D_{3 / 2} ( F = 2) $ state
are made only through off-resonant couplings, so $s_0$ should be almost zero.
In this limit $s_{\tmop{pump}}$ monotonically increases to $s_{\infty}$ which
approaches $1$ for $\Gamma_{\tmop{pump}} \gg \Gamma_{\tmop{loss}}$. This
limiting value is obtained for $t_{\tmop{pump}} \gg \tau_{\tmop{pump}}$ where
$t_{\tmop{pump}}$ is the length of time the pump drive is applied to the ion.

With the 935nm laser's 3GHz EOM off, the losses from the $5 D_{3 / 2} ( F =
2)$ state are due to this state's own lifetime, $\Gamma_{\tmop{decay}} = 1 /
\tau_D \approx 1 / 50 \tmop{ms} = 20 \tmop{Hz}$, and off-resonant couplings to
the $^3 [ 3 / 2]_{1 / 2} (F = 1)$, $\Gamma_{\tmop{off} -
\tmop{resonant}}^{935}$, giving $\Gamma_{\tmop{loss}} = \Gamma_{\tmop{decay}}
+ \Gamma_{\tmop{off} - \tmop{resonant}}^{935}$. In the example shown below
$\Gamma_{\tmop{off} - \tmop{resonant}}^{935}$ is also comparable to
$\Gamma_{\tmop{decay}}$ resulting in $\Gamma_{\tmop{loss}} \approx 50
\tmop{Hz}$.

With normal cooling beams, the pump rate is also due only to off-resonant
couplings to the $6 P_{1 / 2} ( F = 1)$ state, $\Gamma_{\tmop{off} -
\tmop{resonant}}^{370}$. As will be seen below, this rate is about $25
\tmop{Hz}$. This would yield only a relatively small $s_{\infty} \approx 25 /
75 = 1 / 3$. By using the 370nm laser's previously described 2GHz EOM,
$\Gamma_{\tmop{pump}}$ can be increased. This directly drives transitions to
the $6 P_{1 / 2} ( F = 1)$ state from which the ion will decay to either of
the $5 D_{3 / 2}$ hyperfine states with similar probabilities. This enhances
the pump rate, $\Gamma_{\tmop{EOM}}$, by an amount proportional to 2GHz EOM
side-band amplitude so that $\Gamma_{\tmop{pump}} = \Gamma_{\tmop{off} -
\tmop{resonant}}^{370} + \Gamma_{\tmop{EOM}}$

\subsection{Pump Profile}

$\tau_{\tmop{pump}}$, $s_{\tmop{pump}}$, and their dependence on
$t_{\tmop{pump}}$ and the EOM power $P_{\tmop{EOM}}$ can all be determined
with a sequence block like $n_0$ which consists of a pump followed by the
previously described probe sequence. The probe step will yield some average
count $\text{} n ( t_{\tmop{pump}}) = \bar{n}_T - s_{\tmop{pump}}^{} (
t_{\tmop{pump}}) \bar{n}_c $ linearly related to $s_{\tmop{pump}}$ as
described previously. With $s_{\tmop{pump}} ( t_{\tmop{pump}})$ as determined
previously,
\begin{eqnarray*}
  n ( t_{\tmop{pump}}) & = & \bar{n}_T - ( s_{\infty} - ( s_{\infty} - s_0)
  e^{- t_{\tmop{pump}} / \tau_{\tmop{pump}}}_{}) \bar{n}_c\\
  & = & ( \bar{n}_T - s_{\infty} \bar{n}_c) + ( s_{\infty} - s_0) \bar{n}_c
  e^{- t_{\tmop{pump}} / \tau_{\tmop{pump}}}
\end{eqnarray*}
This exponential functional form has three actually independent parameters, so
a fit of $n_{} ( t_{\tmop{pump}})$ in this form directly gives $( \bar{n}_T -
s_{\infty} \bar{n}_c)$, $( s_{\infty} - s_0) \bar{n}_c$ and
$\tau_{\tmop{pump}}$, but can not determine the $ \bar{n}_i$ and $s_i$
independently. As mentioned $s_0 \approx 0$ so it can be eliminated, allowing
$s_{\infty} \bar{n}_c$ to be determined directly. This can then be used to
determine $\bar{n}_T$ as the same product appears in $( \bar{n}_T - s_{\infty}
\bar{n}_c)$.

$\bar{n}_T$ may also be determined directly from an independent measurement
using an $n_1 = \tmop{SPump} / D 2 \tmop{Probe}$ block. $n_1$ is a pump to the
ground state followed by a probe. SPump should give $s_{\tmop{pump}} = 0$ so
that a probe gives $\bar{n}_T$ directly. Ideally a pump step that yields
$s_{\tmop{pump}} = 1$ could be done so that D2Probe would directly give
$\bar{n}_T - \bar{n}_c \equiv \bar{n}_b$ in the same way, and then $\bar{n}_c
= \bar{n}_T - \bar{n}_b$. Such a procedure was not available for these
measurements, though $c_0$ can provide a lower bound on $\bar{n}_b$.

For a single profile, still only the combination $s_{\infty} \bar{n}_c$ is
available. The following examples are profiles that will share the same
$\bar{n}_i$ and $\Gamma_{\tmop{loss}}$, but have varying
$\Gamma_{\tmop{pump}}$. Using $\tau_{\tmop{pump}}$ and $s_{\infty}^{}$ as a
function of $\Gamma_{\tmop{loss}}$ and $\Gamma_{\tmop{pump}}$ relates them in
a way that if a fit is made to all profiles simultaneously, the $\bar{n}_i$,
$s_{\infty}$, $\Gamma_{\tmop{loss}}$ and the various $\Gamma_{\tmop{pump}}$
can be determined independently.

\begin{figure}[h]
  a)\resizebox{0.45\columnwidth}{!}{\includegraphics{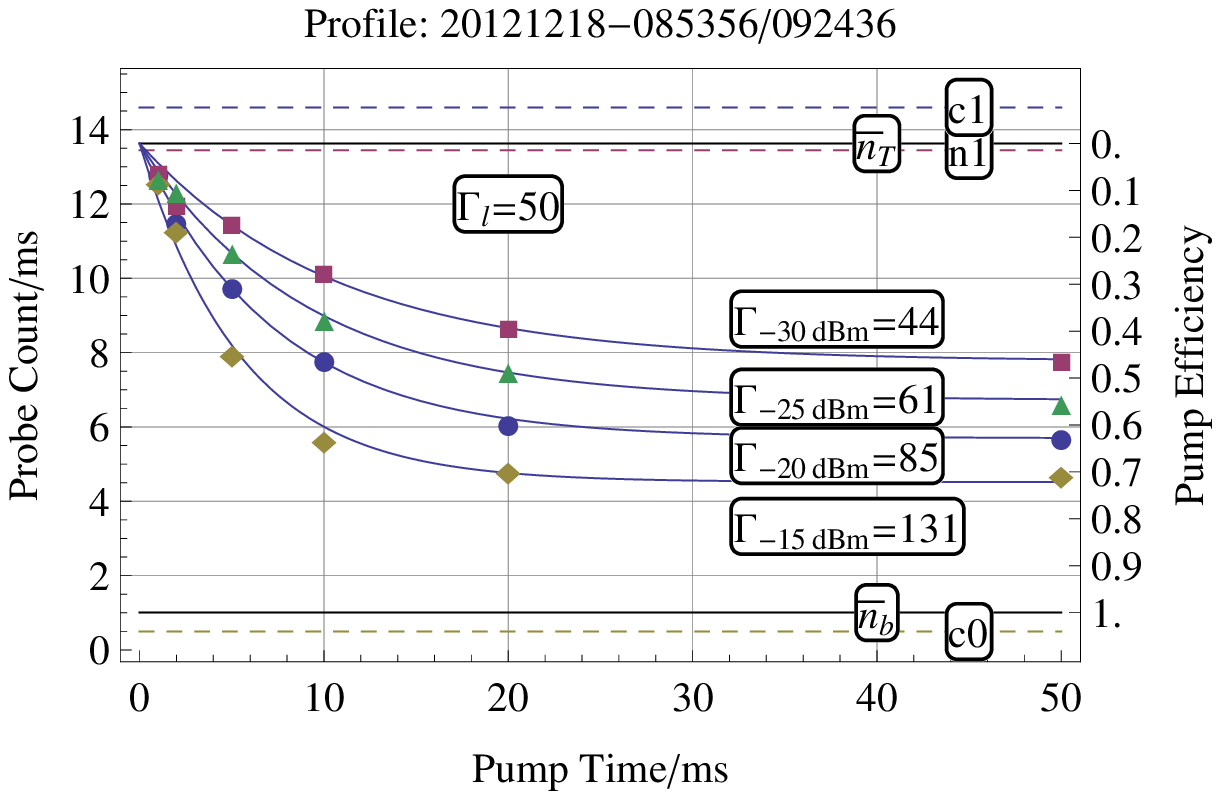}}b)\resizebox{0.40\columnwidth}{!}{\includegraphics{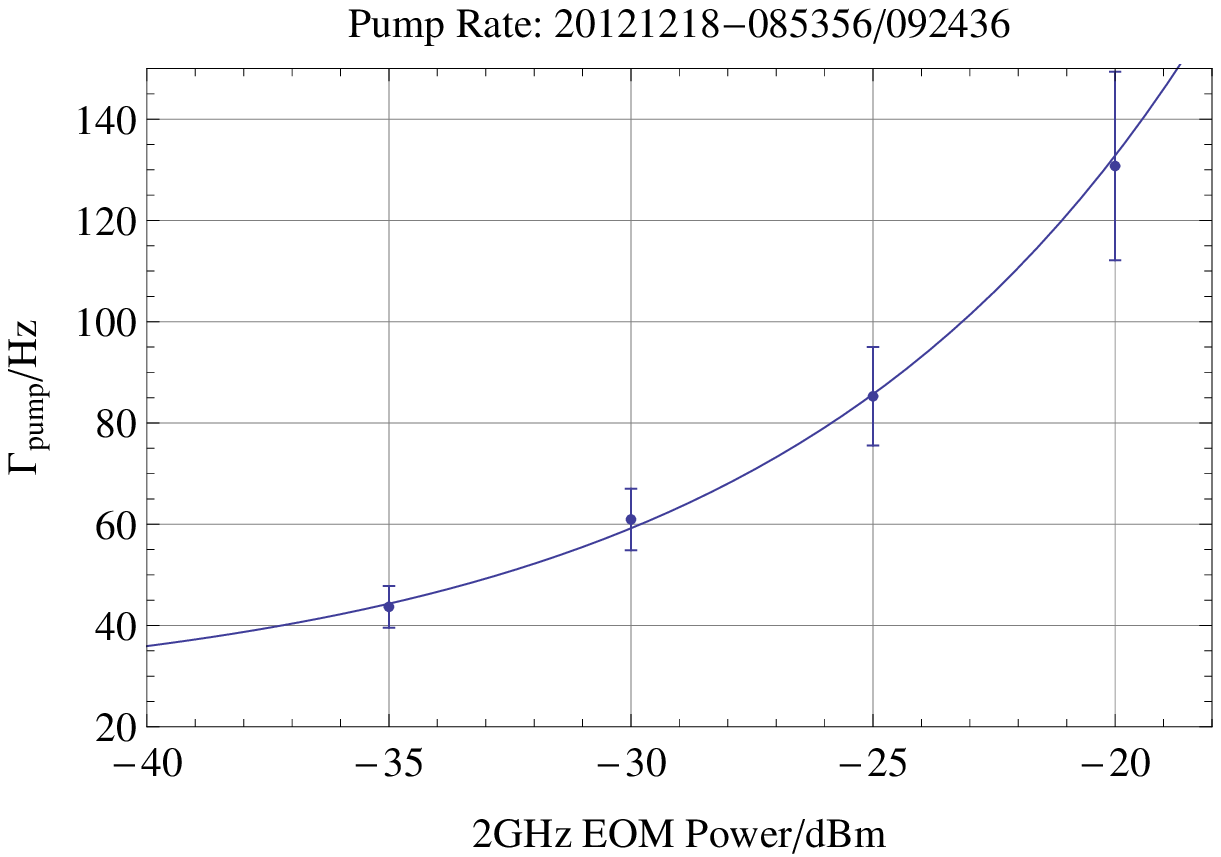}}
  \caption{\label{pump-profile}Number of probe counts as a function of pump
  time for various 2GHz EOM powers, and the resulting pump rate as a function
  of that power. Error bars on the left are determined from the standard
  deviation of the probe counts and are smaller than the plot symbols used.}
\end{figure}

Figure \ref{pump-profile}a shows $n_0$ as a function of $t_{\tmop{pump}}$ for
various 2GHz EOM powers. This data was taken with the same experimental
parameters, most importantly the cooling and background rates, except for the
pump time and EOM oscillator amplitude. An $n_0$ block using each of two EOM
powers is included in each trial and the pump time was changed from trial to
trial to better ensure that possible variations of any other parameters
affected all cases in the same way.

The entire sequence used in this measurement was $c_1$/$n_0 (
t_{\tmop{pump}}, P_{\tmop{EOM}}^{( 1)})$/$n_0 ( t_{\tmop{pump}},
P_{\tmop{EOM}}^{( 2)})$/$n_1$/$c_0$, where $t_{\tmop{probe}}$ was randomly
chosen from the set $\{ 1, 2, 5, 10, 20, 50 \}$ms before each set of trials,
used for 10 trials, and then re-selected for the next set of trials. The time
dependence shows the expected exponential behavior with rates and asymptotic
values determined by $\Gamma_{\tmop{pump}}$ and $\Gamma_{\tmop{loss}}$.

With the instrumentation then at hand, only two EOM powers could be used
during a particular set of trials constituting a single measurement. So two
measurements were made using different pairs of EOM powers, $\{ - 35, - 25 \}
\tmop{dBm}$ and $\{ - 30, - 20 \} \tmop{dBm}$. The stability and consistency
of $n_1$ and $c_1$ between these two runs as seen in the strip chart (not
shown) is a good justification for considering the cooling and probe rates to
be the same between the different sets of EOM power pairs so that the fits can
be made to all profiles simultaneously allowing only $\Gamma_{\tmop{pump}}$ to
change. Thus $\bar{n}_c$, $\bar{n}_b$, $\Gamma_{\tmop{loss}}$ and
$\Gamma_{\tmop{pump}} ( P_{\tmop{EOM}})$, and the derived $s_{\infty}$ and
$\tau$, can all be determined independently. This example exhibits a range of
$\Gamma_{\tmop{pump}}$ from about $40 - 130 \tmop{Hz}$ and
$\Gamma_{\tmop{loss}} \approx 50 \tmop{Hz}$. With $\Gamma_{\tmop{decay}}
\approx 20 \tmop{Hz}$ this provides a measure of $\Gamma_{\tmop{off} -
\tmop{resonant}}^{935} \approx 30 \tmop{Hz}$. Note also that $\bar{n}_T
\approx n_1$, and $\bar{n}_b \gtrsim c_0$ as expected.

\subsection{Pump Efficiency and $P_{\tmop{EOM}}$ Dependence}

With $\bar{n}_b$ and $\bar{n}_c$ determined the actual $s_{\tmop{pump}}$ can
be estimated. These values are used to set the scale on the right-hand axis of
figure \ref{pump-profile}a and show a maximal $s_{\infty}$ of 0.72 for the
highest pump rate. Here optimizing $s_{\tmop{pump}}$ is simply a matter of
maximizing $\Gamma_{\tmop{pump}} / \Gamma_{\tmop{loss}}$ as that both
maximizes $s_{\infty}$ and reduces $\tau_{\tmop{pump}}$ allowing for a shorter
pump time, a shorter trial time and more trials for a given total observation
time $T$. This is mostly a matter of increasing the 2GHz EOM as much as is
possible and practical. Figure \ref{pump-profile}b shows
$\Gamma_{\tmop{pump}}$ as a function of $P_{\tmop{EOM}}$ fitted to a model
that assumes that the $\Gamma_{\tmop{EOM}}$ contribution to
$\Gamma_{\tmop{pump}}$ is proportional to the EOM side-band amplitude
$\Gamma_{\tmop{pump}} = a \sqrt{10^{P_{\tmop{EOM}} / 10}}$ for
$P_{\tmop{EOM}}$ measured in $\tmop{dBm}$, $\Gamma_{\tmop{pump}} =
\Gamma_{\tmop{off} - \tmop{resonant}}^{370} + a 10^{P_{\tmop{EOM}} / 10}$.
This provides the $\Gamma_{\tmop{off} - \tmop{resonant}}^{370} = 25 \tmop{Hz}$
referred to previously.

The total $\Gamma_{\tmop{pump}} \approx 130 \tmop{Hz}$ turns out to be about
the largest possible in practice. Power in the side-bands comes at the expense
of power in the fundamental. When the fundamental power is lowered beyond that
which saturates the cooling transition that overall rate drops and that loss
soon becomes greater than the increases gained from $\Gamma_{\tmop{EOM}}
\nosymbol$. At $P_{\tmop{EOM}} = - 5 \tmop{dBm}$ $s_{\tmop{pump}}$ is already
reduced to 0.5. Subsequent measurements in the $D$ state then use an
approximately optimal $P_{\tmop{EOM}} = - 20 \tmop{dBm} \rightarrow - 15
\tmop{dBm}$. This gives $\tau_{\tmop{pump}} \lesssim 5 \tmop{ms}$ so
$t_{\tmop{pump}} = 15 \rightarrow 20 \tmop{ms}$ is used.

\subsection{$\Gamma_{\tmop{loss}}$ Dependence}

$s_{\tmop{pump}}^{\infty}$ might be further increased by reducing
$\Gamma_{\tmop{off} - \tmop{resonant}}^{935} $ and through it
$\Gamma_{\tmop{loss}}$ by reducing the 935nm laser power during the pump. The
935nm laser is switched by using the first deflected beam after passing the
beam through an acousto-optic modulator (AOM), so the laser power is
essentially controlled by the AOM's RF power input. By applying reduced RF
power to the AOM, the resulting switched laser power is reduced.
$\Gamma_{\tmop{loss}}$ is ultimately limited by $\Gamma_{\tmop{decay}}$, so
this could increase $\Gamma_{\tmop{pump}} / \Gamma_{\tmop{loss}}$ by about a
factor of two. But since we already have $\Gamma_{\tmop{pump}} /
\Gamma_{\tmop{loss}} \gtrsim 2.5$ this would only improve
$s_{\tmop{pump}}^{\infty}$ to about $0.85$. It would also reduce the clean up
rate which may result in a larger probability to be in the $5 D_{3 / 2} ( F =
1)$ level at the end of the probe and negate any gains from increased
$\Gamma_{\tmop{pump}} / \Gamma_{\tmop{loss}}$. In practice there seems to be
no benefit. A sequence $c_1 / n_1 ( P_{935}^{( 1)}) / n_1 ( P_{935}^{( 2)}) /
n_0 / c_0$ was run with $P_{935}^{( 2)} \approx$0.5$P_{935}^{( 1)}$ with no
difference observed between the $n_1$ for both cases. Possibly in this case
$\Gamma_{\tmop{decay}}$ already dominated, or the $5 D_{3 / 2} ( F = 1)$ state
population became significant.

\section{\label{h.qat5quj5m8l8}Probe Efficiency and Sensitivity}

Off-resonant couplings and the finite lifetime of the $5 D_{3 / 2}$ state also
complicate probe sensitivity. A probed ion can be considered to be in either
the shelved state, with probability $s$, or somewhere in the cooling cycle
with the complementary probability $1 - s$. The probe will yield either $n_b$
or $n_T$ counts, respectively, with $n_i = r_i t_{\tmop{probe}}$ where $r_T =
r_c + r_b$, $r_c$ and $r_b$ are the cooling and background rates as before,
and $t_{\tmop{probe}}$ the probe time. The probe time must be long enough that
enough photons are collected that $n_T$ and $n_b$ can be distinguished above
simple Poisson counting statistics.

Ideally the shelved state would be well isolated from the cooling cycle and
long-lived relative to the probe time. In this case $t_{\tmop{probe}}$ can be
made arbitrarily long, and $n_i$ can be made a large as necessary to allow the
initial state to be determined with almost perfect accuracy.

\subsection{Non-ideal Couplings and Finite Probe Time}

For the $^{171} \tmop{Yb}^+$ $5 D_{3 / 2} ( F = 2_{})$ state, the same
$\Gamma_{\tmop{off} - \tmop{resonant}}^{370}$ contributing to
$\Gamma_{\tmop{pump}}$ will result in the ion being driven to the shelved
state after long probe times thereby reducing the total number of probe
photons collected for an ion beginning in the $S$ state resulting in $n_T
\rightarrow \bar{n}_T < r_T t_{\tmop{probe}}$. Similarly
$\Gamma_{\tmop{loss}}$ from $\Gamma_{\tmop{decay}}$ and $\Gamma_{\tmop{off} -
\tmop{resonant}}^{935}$ result in an ion initially in the $D$ state to be
driven to the $S$ state resulting in $n_b \rightarrow \bar{n}_b > r_b
t_{\tmop{probe}}$. The probe beams effectively become a set of pump beams so
that $s$ is determined by the exponential profile given before as a function
of $t_{\tmop{probe}}$, but in this case with a time scale $\tau_{\tmop{probe}}
= 1 / \Gamma_{\tmop{probe}} \nocomma$ and $\Gamma_{\tmop{probe}}$ now given by
$\Gamma_{\tmop{pump}} = \Gamma^{370}_{\tmop{off} - \tmop{resonant}}$ and
$\Gamma_{\tmop{loss}} =^{} \Gamma_{\tmop{off} - \tmop{resonant}}^{935} +
\Gamma_{\tmop{decay}}$. For very long probe times, $t_{\tmop{probe}} \gg
\tau_{\tmop{probe}}$, the ion's probability to be in the $D$ state is given by
the corresponding steady state value $s_{\infty}^{\tmop{probe}} =
\Gamma_{\tmop{pump}} / ( \Gamma_{\tmop{pump}} + \Gamma_{\tmop{loss}})$
yielding a probe count of $n \approx n_T - s_{\tmop{probe}}^{\infty} n_c
\nosymbol$ independent of the initial state of the ion.

All the information about the initial state of the ion was gathered during
the beginning of the probe, during $t < \tau_{\tmop{probe}}$. Beyond this
there will be only a fixed number of extra counts, $\Delta n$, for an ion
beginning in the $S$ state compared to beginning in the $D$ state, independent
of the probe time. Increasing the probe time increases the counts collected,
but not $\Delta n$. For $t > \tau_{\tmop{probe}}$, as the counts increase so
do the fluctuations in that count, $\sigma_n$. At some point $\sigma_n >
\Delta n$ and the information about the initial state is effectively lost, or
at best not improved while further increasing the probe time only increases
the trial time and reduces the number of trials that can be made.

$\tau_{\tmop{probe}}$ is then effectively a probe coherence time. The actual
probe time $t_{\tmop{probe}}$ must be made less than $\tau_{\tmop{probe}}$ for
the probe to be sensitive to the initial state. This limits the ability to
determine the state of an ion using a single trial. For the characteristic
rates seen in the previous pumping data $\Gamma_{\tmop{probe}} \approx 75
\tmop{Hz}$ so that $\tau_{\tmop{probe}} \approx 10 - 15 \tmop{ms}$. For
cooling rates around 2-4kcps, and background rates 500cps, a probe time of
$t_{\tmop{probe}} = 5 \tmop{ms}$ would give $n_c = 15$ and $n_b = 5$. Poisson
distributions with these means overlap considerably and in real experiments
fluctuating rates, and especially multiple ions, further obscure the
difference.

\subsection{Probe Count Distribution}

\begin{figure}[h]
  a)\resizebox{0.35\columnwidth}{!}{\includegraphics{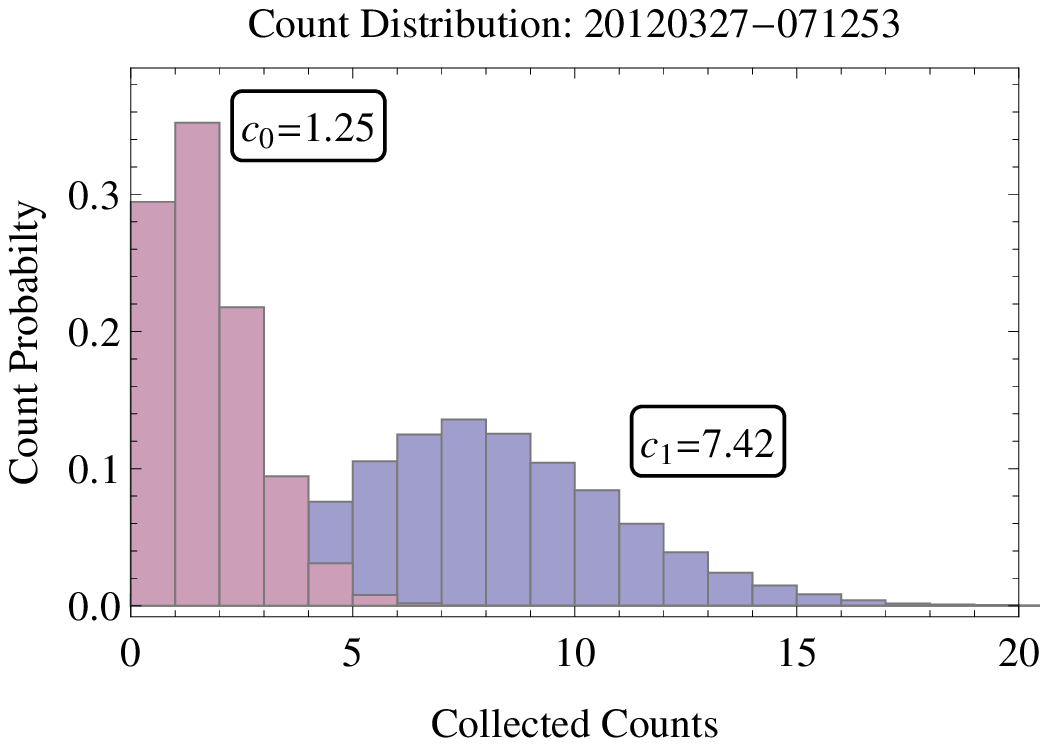}}b)\resizebox{0.55\columnwidth}{!}{\includegraphics{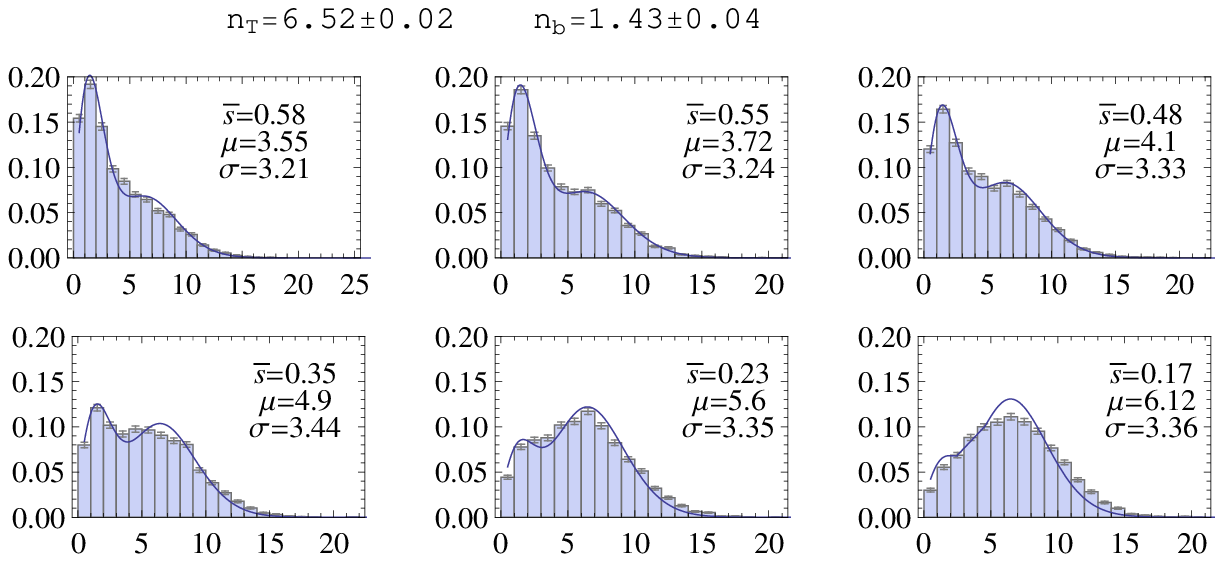}}
  
  c)\resizebox{0.45\columnwidth}{!}{\includegraphics{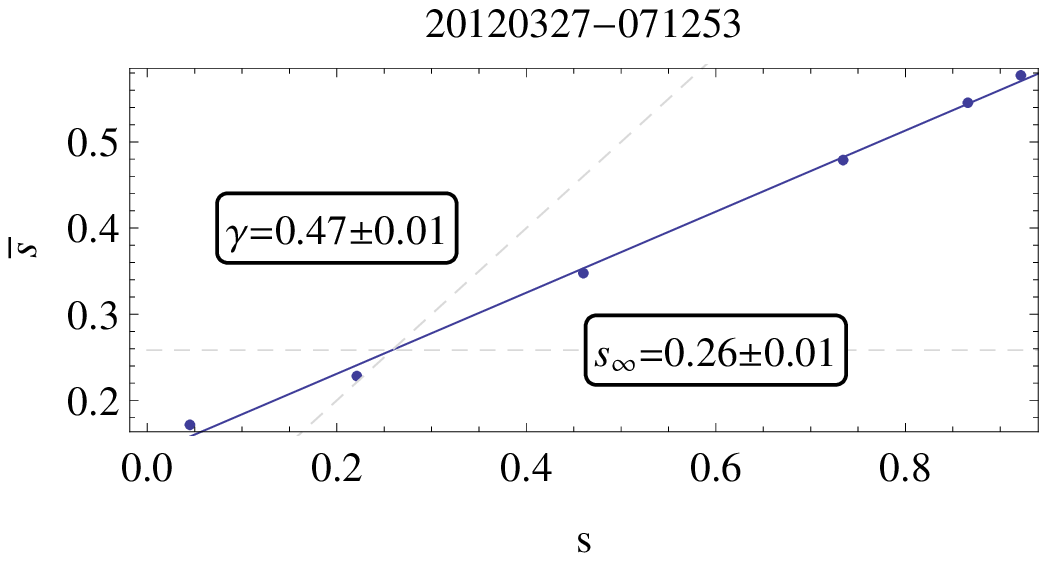}}d)\resizebox{0.45\columnwidth}{!}{\includegraphics{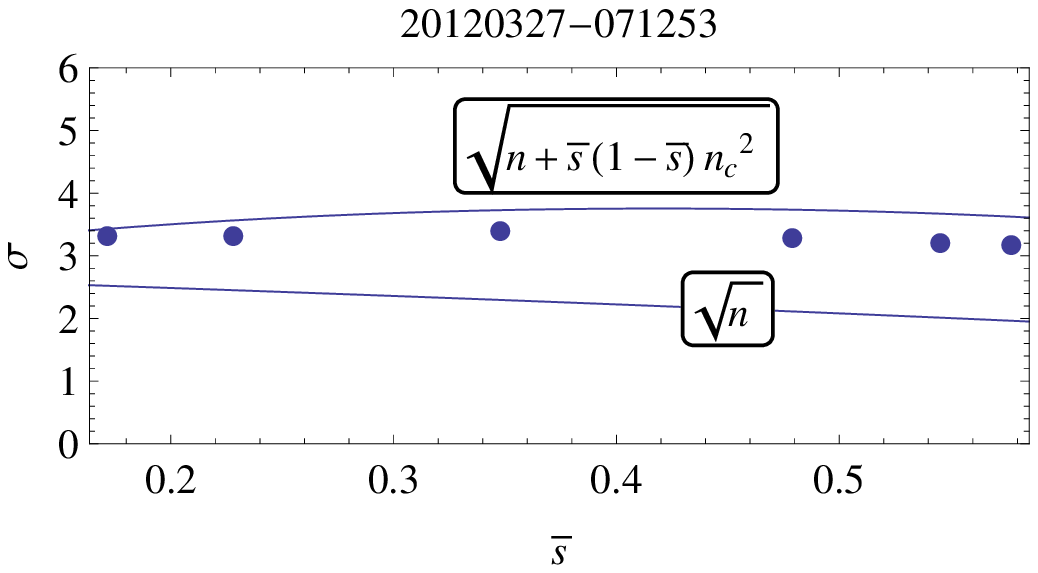}}
  \caption{\label{probe-histogram}Histograms of the counts collected for
  various probes. a) shows the cooling and background rates and the simple
  Poisson distributions that they are expected to follow with parameters
  determined directly from the mean of the distributions. b) are distributions
  from D2Probe with different initial probabilities $s$ to be in the $5 D_{3 /
  2} ( F = 2)$ state. The solid lines are from the expected bi-modal
  distributions with parameters determined by fits to the histograms. $c_1$
  and $c_0$ are shared between all, and $\bar{s}$ is allowed to vary for each
  case. c) shows $\bar{s} ( s)$ using the $\bar{s}$ determined from the
  histogram fits and $s$ from a fit to the lifetime profile that this data was
  intended to measure. d) shows the actual variation of the probe counts for
  each $\bar{s}$ as determined from the standard deviation of the probe counts
  along with that expected for both a Poisson distribution, and a bi-modal
  distribution each with parameters as determined from the data. The
  uncertainties of the widths, as determined from the data are smaller than
  the plot points.}
\end{figure}

Figure \ref{probe-histogram} shows typical histograms of $c_1$, $c_0$, and the
$n_0$ resulting from a probe for various $s$ for a single ion. The first two
are described very accurately by the expected Poisson distribution, indicating
a stable cooling rate. The remaining should follow a bi-modal
distribution[\ref{schacht-shelving}] given by,
\begin{eqnarray*}
  P_n ( s) & = & ( 1 - \bar{s}) p_n ( n_T) + \bar{s} p_n ( n_b)
\end{eqnarray*}
where
\begin{eqnarray*}
  \bar{s} & = & \gamma ( s - s_{\infty}) + s_{\infty}\\
  \gamma & = & \frac{1 - e^{- t_{\tmop{probe}} /
  \tau_{\tmop{probe}}}}{t_{\tmop{probe}} / \tau_{\tmop{probe}}}
\end{eqnarray*}
$p_n ( n_i)$ is a Poisson distribution with mean $n_i$, and $s_{\infty}$
corresponds to that given by the probe beams. Note that for $t_{\tmop{probe}}
/ \tau_{\tmop{probe}} \rightarrow 0$, $\gamma \rightarrow 1$ and $\bar{s}
\rightarrow s$. This explicitly includes the effects of all the non-ideal
couplings that have been discussed. The data used here is from one set that is
later used to determine the $D$ state lifetime so the various $s$ are
generated by a variety of wait times after pumping. $P_n ( \bar{s})$ is fit to
all profiles collectively using a common $n_c$ and $n_b$ with a profile
dependent $\bar{s}_i$. $n_T$ and $n_b$ should be given by $c_1 = 7.4$ and $c_0
= 1.3$ respectively. The fits yield $n_b = 1.4$ but a slightly lower $n_T =
6.5$.

The models reproduce most of the general features of all the distributions
including the variances. For a Poisson-distributed random variable the
variance is given simply by $\sigma^2 = n$. For this single ion bi-modal
distribution there is an additional contribution[\ref{schacht-shelving}],
\begin{eqnarray*}
  \sigma^2 & = & n + \bar{s} ( 1 - \bar{s}) n_c^2
\end{eqnarray*}
Figure \ref{probe-histogram}d shows the measured widths, computed from the
standard deviation of the data, as a function of $\bar{s}$. The variances are
clearly larger than for a Poisson distribution. To compare to the bi-modal
distribution width, uncertainties in the width are estimated by regarding the
standard deviation as any other function and propagating fluctuations in $n$
through, giving simply $\sigma_{\sigma} = \sigma / \sqrt{N_{\tmop{trials}}}$.
The predicted error also generally follows the measured standard deviation but
misses the precise details.

\begin{figure}[h]
  \resizebox{0.5\columnwidth}{!}{\includegraphics{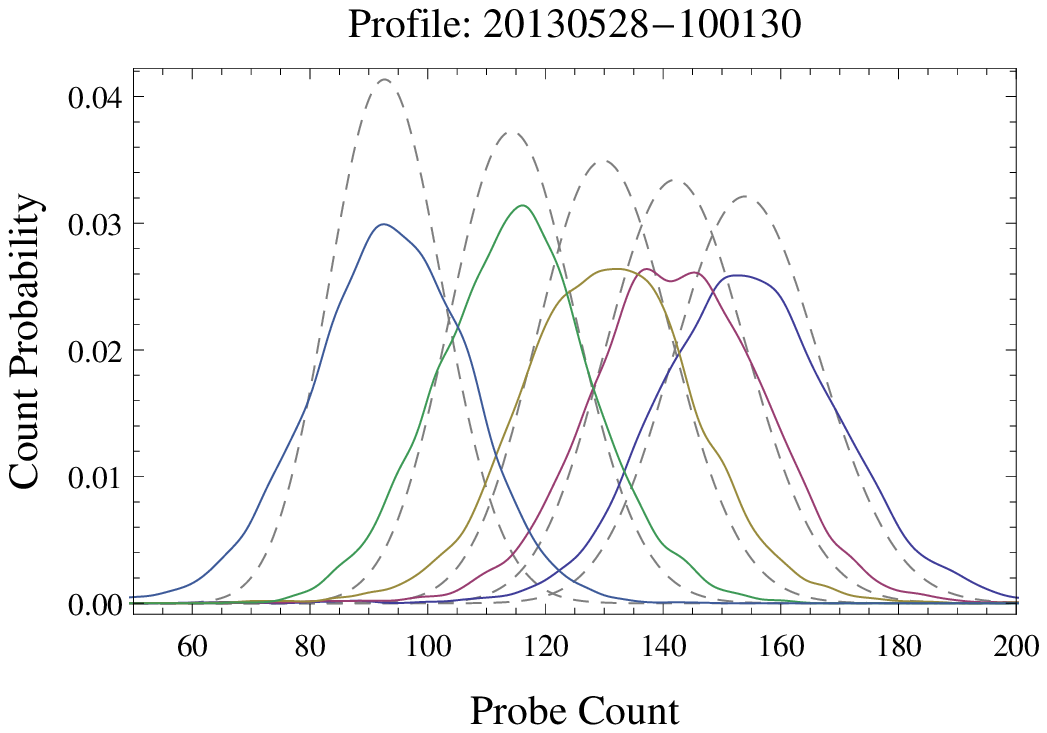}}\resizebox{0.45\columnwidth}{!}{\includegraphics{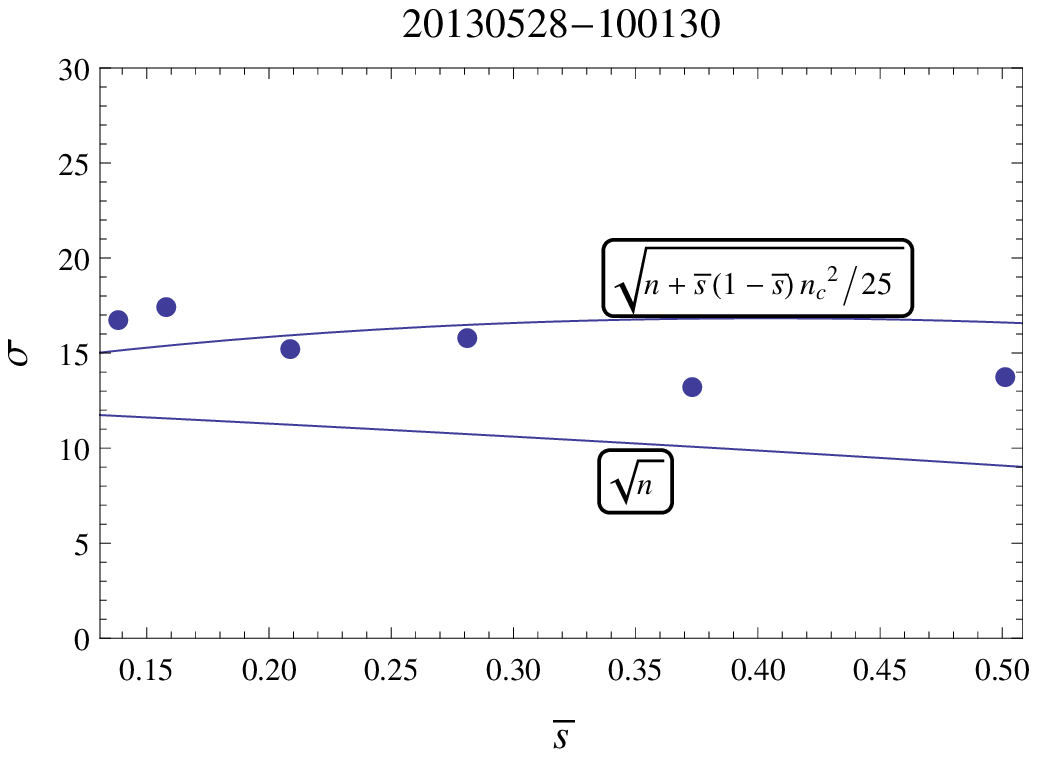}}
  \caption{\label{large-N-distribution}Probe count distribution and variation
  for various $\bar{s}$ for a large number of ions. The solid curves on the
  left are the actual distribution, and the dashed curves the Poisson
  distributions they are expected to approach in the $n \rightarrow \infty$
  limit.}
\end{figure}

For $N_{\tmop{ions}} > 1$ it turns out[\ref{schacht-shelving}] that
\begin{eqnarray*}
  \sigma^2 & = & n + \frac{\bar{s} ( 1 - \bar{s})}{N_{\tmop{ions}}} n^2_c
\end{eqnarray*}
and the distribution gradually converges to a Poisson. Figure
\ref{large-N-distribution} shows a similar series of probe distributions for
various $s$ for cases with $N_{\tmop{ions}} = 20 - 30$ ions with the Poisson
distribution that corresponds to their means. The distributions are clearly
narrower, though in this case still slightly larger than a Poisson. The
prediction for the width is not as accurate as before. This is likely mostly
due to an inaccurate estimate of $n_T$ or $n_b$ and the resulting $\bar{s}$,
or the number of ions.

The remaining differences between the observations and predicted widths and
distributions for this and the $N = 1$ differences might be due to probe rate
fluctuations, but a fit using a convolution of $p_n$ over a uniformly
distributed variation of $n_c$ over $\pm 20$\% does not qualitatively improve
the results and that range is already more than can be justified in light of
the stability of $c_1$. The discrepancies may also be due to the assumption of
simple first order rate equations being invalid due to non-negligible
populations in disregarded intermediate states. Further attempts to account
for these disparities have not yet been made.

For present purposes note only that the probe count histograms indicate that
there is no well-defined distinction between the results of probing a shelved
or un-shelved ion, even for a single ion, so the results of a single probe can
not be used to determine the initial state. Again, the mean is used instead,
\begin{eqnarray*}
  n ( t_{\tmop{probe}}) & \equiv & \langle n \rangle_{P_n ( s)} = ( 1 -
  \bar{s}) n_T + \bar{s} n_b = n_T - \bar{s} n_c
\end{eqnarray*}
Which can be written$\begin{array}{lll}
  n ( t_{\tmop{probe}}) & = & \bar{n}_T - s \bar{n}_c
\end{array}$ as before with
\begin{eqnarray*}
  \bar{n}_b & = & \nobracket n_b - n_c ( 1 - s^{\infty}_{\tmop{probe}}) ( 1 -
  \gamma \nobracket)\\
  \bar{n}_c & = & n_c \gamma
\end{eqnarray*}
Note that $\gamma = \bar{n}_c / n_c$ is a direct proportional measure of how
much of the difference between cooling and background rates is traversed by
$n$ over the range of $0 < s < 1$. This makes $\gamma$ a measure of probe
efficiency. For $t_{\tmop{probe}} / \tau_{\tmop{probe}} \rightarrow 0$ these
reduce to, $\bar{n}_i \rightarrow n_i$. For $t_{\tmop{probe}} \gg \tau$,
$\bar{n}_c \rightarrow 0$ and both $\bar{n}_T, \bar{n}_b \rightarrow n_b - n_c
( 1 - s^{\infty}_{\tmop{probe}})$ and the resulting probe count becomes
insensitive to the initial state.

$n ( t_{\tmop{probe}})$ is a linear function of $s$ and so can be used to
reliably determine at least the dependence of $s$ on the wait time
$t_{\tmop{wait}}$ for this measurement, and even $s$ itself if $\bar{n}_c$ and
$\overline{n_{}}_b$ can also be determined. For the lifetime data used in the
previous histograms $\bar{n}_T$ can be determined from the decay profile to be
$\bar{n}_T = 6.25 \pm 0.02$. As usual, $\bar{n}_b$ can not be determined
directly, but it might be estimated by $\bar{n}_b \approx c_0 + ( c_1 -
\bar{n}_T)$. The $s_{t_{\tmop{wait}}}$ can then be determined from
$s_{t_{\tmop{wait}}} = ( \bar{n}_T - n_{t_{\tmop{wait}}}) / \bar{n}_c$. The
$\bar{s}_i$ are determined from the histogram fits and the resulting $\bar{s}
( s)$ fit to get $\gamma$ and $s_{\tmop{pump}}^{\infty}$. These can then be
used to get a new estimate for $\bar{n}_b$ and the fit repeated. This
procedure quickly converges to $\bar{n}_b = 2.54$, $\gamma = 0.60$,
$s_{\infty} = 0.34$. The resulting final $\bar{s} ( s)$ and fit are shown in
figure \ref{probe-histogram}c. A direct calculation from the $\bar{n}_{_i}$
and $n_i$, give $\gamma = \bar{n}_c / n_c = 0.73$, $s_{\tmop{pump}}^{\infty} =
0.20$ which should be, but are not particularly consistent with the values
obtained from the fit to $\bar{s} ( s)$, suggesting some further lack of
understanding. Either result gives $\tau_{\tmop{probe}} \approx 1.2
t_{\tmop{probe}}$. Here the probe time was $5 \tmop{ms}$ so that
$\tau_{\tmop{probe}} \approx 6 \tmop{ms}$.

\subsection{Probe Profile and Sensitivity}

\begin{figure}[h]
  \resizebox{0.65\columnwidth}{!}{\includegraphics{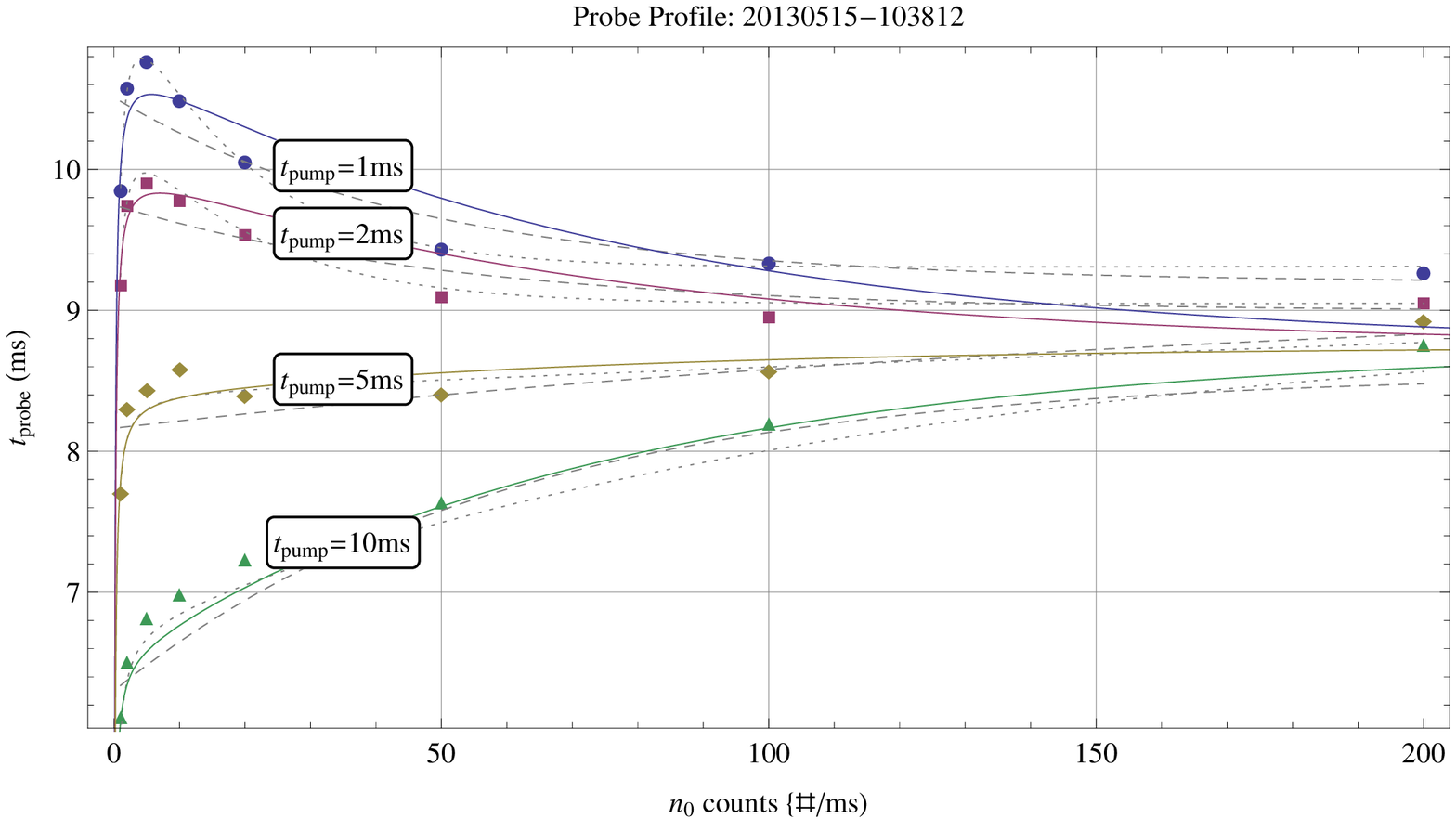}}\begin{tabular}{c}
    
  \end{tabular}\resizebox{0.25\columnwidth}{!}{\includegraphics{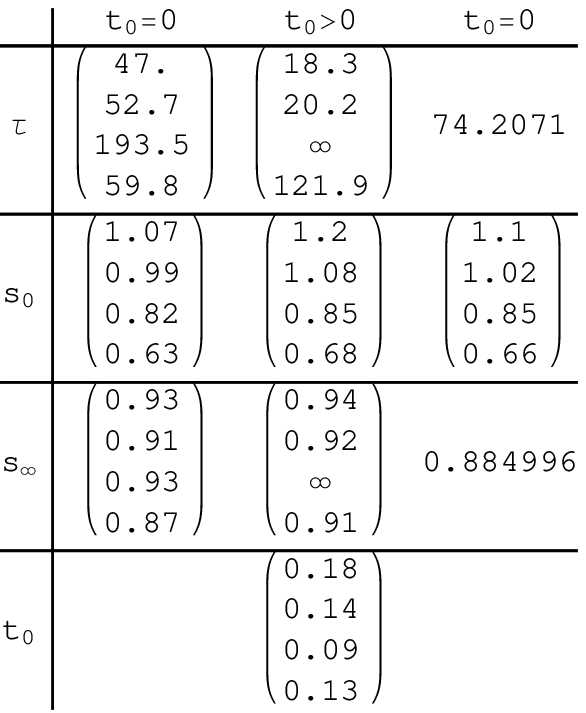}}
  \caption{\label{probe-profile}Average probe counts per time as a function of
  probe time for various initial $s$. The dashed lines are individual fits to
  each profile assuming $t_0 = 0$, the dotted lines are the same but allowing
  $t_0 \neq 0$. The solid lines are the best fit to all profile
  simultaneously, taking $t_0 = 0$, all sharing the same $\tau$ and
  $s_{\tmop{probe}}^{\infty}$. The table shows the fit parameters
  corresponding to each case. }
\end{figure}

With the probe distribution and mean understood the $t_{\tmop{probe}}$
dependence of the parameters can be investigated. Figure \ref{probe-profile}
shows the results of $n_0 ( t_{\tmop{probe}})$ for various initial $s$
generated by varying the pump times so that $s = s_{\tmop{pump}} (
t_{\tmop{pump}})$. The complete sequence is $c_1 / c_{1 a} / n_1 (
t_{\tmop{probe}}) / n_0 ( t_{\tmop{pump}}, t_{\tmop{probe}}) / c_0$ where as
usual the times are chosen randomly from $t_{\tmop{pump}} = \{ 1, 2, 5, 10 \}
\tmop{ms}$ and $t_{\tmop{probe}} = \{ 1, 2, 5, 10, 20, 50, 100, 200 \}
\tmop{ms}$. Three fits are shown, the first is an individual fit to the
profile for each probe time to $n$ as a function of $t_{\tmop{probe}}$ where
the $\bar{n}_i$ are now regarded as functions of $t_{\tmop{probe}}$
\[ n ( t_{\tmop{probe}}) = \bar{n}_T ( t_{\tmop{probe}}) - s^{}  \bar{n}_c (
   t_{\tmop{probe}}) \]
This is implicitly a function of $n_i$, $s_0$, $s_{\infty}$ and
$\tau_{\tmop{probe}}$, but the form again reduces to a three parameter
exponential. $c_1$ and $c_0$ can be used to determine $n_T$ and $n_b$
respectively, and the remaining quantities derived from these and the fit
parameters.

These fits are qualitatively good, but still show apparent statistically
significant systematic deviations from the expected behavior, especially for
short probe times. The resulting fit parameters are fairly consistent giving
$s_{\infty} \approx 0.9$, \ $\bar{n}_b \gtrsim c_0$, and $\tau_{\tmop{probe}}
= 45 - 60 \tmop{ms}$ approximately less then $\tau_{\tmop{decay}}$, though not
by as much as expected, and with the exception of $\tau_{10 \tmop{ms}} \approx
200 \tmop{ms}$.

The short time fit can be improved by including one more imperfection. The
fit data are the effective rates, the probe counts divided by the probe time,
rather than the probe counts directly. This requires that the actual real
probe time is known. While the gates that direct the beams states can be timed
to better than $\mu s$, shutter lags and gate delays can result in the
actually exposure time during the probe to be different. Suppose that the
difference is $t_0$ independent of $t_{\tmop{probe}}$. Then the counts should
be normalized by $t_{\tmop{probe}} + t_0$ rather than just $t_0$, so the
effective count rate is wrong by a factor $t_{\tmop{probe}} / (
t_{\tmop{probe}} + t_0)$. This correction should be included with
$t_{\tmop{probe}}$ in the exponentials that appear in $\gamma$ as well, though
they will just effectively modify $\bar{n}_c$ and $\bar{n}_b$ and not give
qualitatively different functional behavior.

Alternately the $\bar{n}_T$ as measured by $c_1$ may not be accurate. $c_1$
is measured during a cooling period when all cooling beams have been on for a
while, while the $n_0 $ probe starts with all beams off and then the beam
blocks and PMT gates are all switched simultaneously. Again, incorrectly
calibrated shutter lags may result in the beams being on shorter, or longer,
than intended so that $c_1$ is not equal to the maximum possible $n_1$ by a
similar factor. In this case $c_1$ and $c_0$ should be modified by this
factor, but not $t_{\tmop{probe}}$. The result is the same functional form,
but the resulting $s_i$ will be slightly different. In this case $t_0$ turns
out to be small enough that these choices are indistinguishable.

With this modification most of the short pump probe time behavior matches
precisely with a reasonable $t_0 \approx 0.1 \tmop{ms}$, though there are
still statistically significant variations from the model for longer probe
times, and the fit parameters are even more inconsistent. Two of the $\tau$
yield a much closer to expectations $20 \tmop{ms}$, but the two others are
unreasonably large. Similarly $s_{\tmop{probe}}^{\infty} \approx 0.9$ except
for $t_{\tmop{pump}} = 5 \tmop{ms}$ where it is greater than 1.

A collective fit to all profiles simultaneously allowing only $s$ to vary with
$t_{\tmop{probe}}$, makes matters worse. In this case the fit is poor for
almost all probe times, and $\tau_{\tmop{probe}}$ is an unreasonably long $80
\tmop{ms}$. A pump time analysis could be done to independently determine
$\Gamma_{\tmop{loss}}$ and $\Gamma_{\tmop{pump}}$ and compare the resulting
$s_{\tmop{pump}}$ to the $s$ obtained from the probe time fits. That can't be
done with this particular data though it was taken with different pump times,
because the pump stage uses the $2 \tmop{GHz}$ EOM and a full power $935
\tmop{nm}$ laser, while the probe uses the complement, thus providing
$\Gamma_{\tmop{pump}} = \Gamma_{\tmop{off} - \tmop{resonant}}^{370} +
\Gamma_{\tmop{EOM}}$, and $\Gamma_{\tmop{loss}} = \Gamma_{\tmop{decay}} +
\Gamma_{\tmop{off} - \tmop{resonant}}^{935}$ rather than the
$\Gamma_{\tmop{pump}} = \Gamma_{\tmop{off} - \tmop{resonant}}^{370}$, and
$\Gamma_{\tmop{loss}} = \Gamma_{\tmop{decay}} + \Gamma_{\tmop{off} -
\tmop{resonant}}^{935 \tmop{attenuated}}$ required here. Similarly the $n_0 (
t_{\tmop{pump}}, t_{\tmop{probe}})$ count distributions could be used to
determine $s$ in each case.

Such efforts may highlight an important discrepancy, but the results would
not yield a better fit and the model's functional dependence is clearly wrong.
It is not immediately clear how to account for the disparity. Non-negligible
intermediate state populations may again be a factor. Such consideration have
not yet been pursued as they are becoming too far removed from the primary
purpose of these measurements. Though this does not allow for a completely
rigorous justification for assuming that $n_0 ( s)$ is precisely linear. A
more complicated non-linear behavior might lead to some systematic variation
of a parameter derived from fits to $n_0$, though the fairly precise
exponential form of the pump profile and the very precise exponential form of
the decay profiles seen later are a compelling empirical justification. It is
unlikely that non-linear correction to $n_0 ( s)$ would combine with $s (
t_{\tmop{wait}})$ or $s ( t_{\tmop{pump}})$ to result in an identical
functional form.

\subsection{Optimal Probe}

Though the model doesn't precisely predict the details, the data is sufficient
to confirm an approximate probe coherence time of $\tau_p \approx 10 - 20
\tmop{ms}$. As discussed previously, a probe time much longer than this loses
sensitivity because the probe count rate becomes independent of the initial
state of the ion. Conversely, a short probe time has poor sensitivity because
of low counting statistics. An intermediate time will minimize the uncertainty
in determining $s$. The $\sigma_n$ given by the kind of multi-modal count
distributions exhibited here turns out to result in an approximately minimal
$\sigma_s$ at $t_{\tmop{probe}} \approx \tau_{\tmop{probe}} / 2$
[\ref{schacht-shelving}]. With the preceding determination of
$\tau_{\tmop{probe}} \approx 10 - 15 \tmop{ms}$, an optimal probe for this
system would have $t_p = 5 - 10 \tmop{ms}$. At this point $\sigma_s$ is given
by
\begin{eqnarray*}
  \sigma_{N s} & = & \sigma_s \sqrt{\frac{t_0}{T}}\\
  \sigma_s & \approx & 1.27 ( 2 a_{\max} + 1)^{1 / 2} / \sqrt{4
  N_{\tmop{ions}}}\\
  a_{\max} & = & \frac{r_T}{r_c^2 \tau_{\tmop{probe}}} 4 N_{\tmop{ions}}
\end{eqnarray*}

For the parameters determined in the previous example $r_T \approx 6.52 /
\tmop{ms}$, $r_b \approx 1.43 / \tmop{ms}$, give $a_{\max} \approx 0.05$
corresponding to fairly good counting statistics. For $t_0 \approx 135
\tmop{ms}$ this gives $\sqrt{t_0} = 6 \times 10^{- 3} \sqrt{\tmop{hr}}^{}$,
$\sigma_s = 0 .66$ and an overall sensitivity of order
\begin{eqnarray*}
  \sigma_{N s} & \sim & 4 \times 10^{- 3} / \sqrt{T / \tmop{hr}}
\end{eqnarray*}
about 0.4\%$\sqrt{\tmop{hr}}$.

\subsection{935nm Laser Power}

$\tau_{\tmop{probe}}$ depends on the 935nm laser power through
$\Gamma_{\tmop{loss}} = \Gamma_{\tmop{decay}} + \Gamma_{\tmop{off} -
\tmop{resonant}}^{935}$. Data presented previously indicated
$\Gamma_{\tmop{off} - \tmop{resonant}} \sim 20 \rightarrow 30 \tmop{Hz}$.
$\tau_{\tmop{probe}}$ could be increased, and $\sigma_s$ reduced, by reducing
$\Gamma_{\tmop{off} - \tmop{resonant}}^{935}$ with reduced 935nm laser power.
This will improve the sensitivity as long as the resonant 935nm transition
remains saturated. If 935nm intensity is reduced below saturation then $r_c$
drops as well, but linearly compared to $\sqrt{\tau_{\tmop{probe}}}$ which
then increases $\sigma_s \nosymbol$. When a further decreasing
$\Gamma_{\tmop{off} - \tmop{resonant}}$ then becomes much less than
$\Gamma_{\tmop{decay}}$, $\tau_{\tmop{probe}}$ doesn't increase as quickly and
$\sigma_s$ increases even more quickly with lower 935nm laser power. So gains
in sensitivity may be realized by reducing 935nm laser power to the point that
the resonant transition is just saturated, though that gain may be modest if
$\Gamma_{\tmop{loss}}$ is already dominated by $\Gamma_{\tmop{decay}}
\nosymbol$.

\begin{figure}[h]
  \resizebox{0.95\columnwidth}{!}{\includegraphics{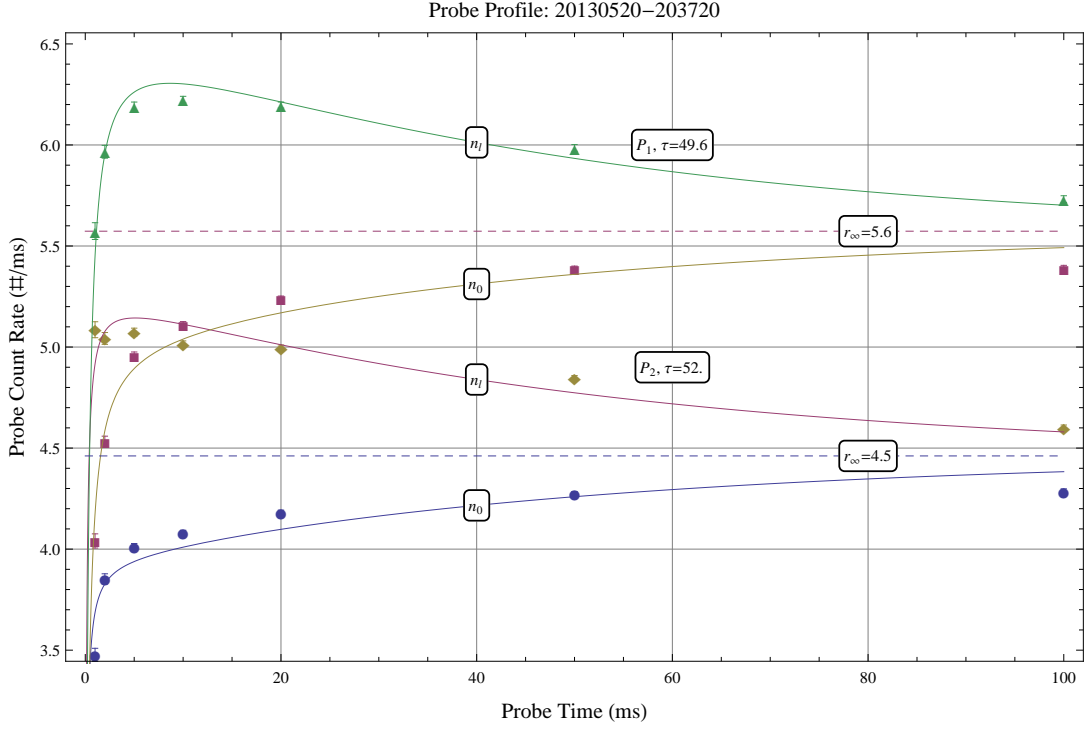}}
  \caption{\label{alt-probe}Probe profile for two different 935nm laser
  powers, $P_1 \approx 2 P_2$.}
\end{figure}

As when discussing pump efficiency, the 935nm laser power can be controlled
by changing the switch AOM's RF drive amplitude. Figure \ref{alt-probe} shows
the results of a sequence that includes the block $n_0 ( P_1^{935})$ and $n_0
( P_2^{935})$, with $P_2^{935} \approx 0.5 P_1^{935}$. The probe using the
higher power does indicate a slightly shorter $\tau_{\tmop{probe}}$, but also
a slight higher $r_c$ so that in this case there is little difference in
sensitivity between the two.

In other cases modest improvements have been seen and so are sometimes used.
Generally the reduced power is set by observing the cooling rate and reducing
935nm laser power until this rate just starts to decrease. When the reduced
probe rate is used a $c_{1 a}$ block is also included which measures the
cooling rate with the reduced 935nm laser power to provide $n_c$. In all the
examples presented here, when this is the case, it is implied that $c_1$
refers to this attenuated result.

\subsection{General Considerations}

The most important factor affecting probe sensitivity is
$\tau_{\tmop{probe}}$. Using $t_{\tmop{probe}}$ much different than
$\tau_{\tmop{probe}} / 2$ significantly reduces
sensitivity.[\ref{schacht-shelving}] The rates that determine
$\tau_{\tmop{probe}}$ are very stable during a particular run, but are not
easily controlled and require some work to determine. They are always of the
same general size, so for new experiments typical values of $t_{\tmop{probe}}
= 10 - 15 \tmop{ms}$ are used and rates are only remeasured to provide better
$t_{\tmop{probe}}$ if $\gamma$ becomes less than about $0.3$. $\gamma$ can be
determined from $\bar{n}_c / n_c$. $\bar{n}_T$, $n_c$, and $n_b$ can all be
determined directly from $n_1$, $c_1$ and $c_0$ respectively, but $\bar{n}_b$
can't currently be measured directly due to the difficulty in preparing a
state with $s = 1$ using the pumping procedure discussed, but it should be
about as close to $c_0$ as $\bar{n}_c$ is to $c_1$, depending on what
$s_{\infty}^{}$ for the probe turns out to be. So knowing $n_i$ and $c_i$
provides enough information to estimate $\gamma$.

There remain some uncertainties about the details of the probe dynamics, in
particular the profile of $n_0$ as a function of $t_{\tmop{probe}}$, and its
distribution and resulting $\bar{s} ( s)$, each of which fails to precisely
match the statistical model, but for a fixed probe time it appears justified
to consider the results of a probe to be very accurately linearly related to
$s$.

\section{$\label{h.s47d30vs4idr}^{171} \tmop{Yb}^+$ $5 D_{3 / 2} (F = 2)$
Lifetime}

With these details about the pump and probe steps determined and optimized, a
lifetime measurement becomes completely straightforward. After the pump step
the ion is in the $5 D_{3 / 2} ( F = 2)$ state with probability
$s_{\tmop{pump}}$. At time $t_{\tmop{wait}}$ after the end of the pump step
that probability becomes$s_{\tmop{pump}} e^{- t_{\tmop{wait}} / \tau_D}$. A
probe then yields an average count of
\begin{eqnarray*}
  n ( t_{\tmop{wait}}) & = & \bar{n}_T - s_{\tmop{pump}} \bar{n}_c e^{-
  t_{\tmop{wait}} / \tau_D}\\
  & = & \bar{n}_T - \bar{n}_c' e^{- t_{\tmop{wait}} / \tau_D}
\end{eqnarray*}
The sequence block $d_0 = D 2 \tmop{Pump} / \tmop{Off} ( t_{\tmop{wait}}) / D
2 \tmop{Probe}$ yields this particular $n ( t_{\tmop{wait}})$.

\subsection{Sensitivity and Optimal Sampling}

Pump and probe parameters can be chosen to minimize $\sigma_s$, but the
uncertainty, $\sigma_{\tau}$, in determining $\tau$ from the results of $n (
t_{\tmop{wait}})$, can also be affected by the choice of wait times $t_i$ and
the relative frequencies with which they are used. Some sets of wait times can
give an $n$ that is more sensitive to $\tau_D$ than others. Since the
$\bar{n}_i$ cannot be determined independently, they must be also determined
from the data along with the lifetime. There are three effective parameters in
$n$ so data from trials with at least three different wait times must be used.

The uncertainty in the $\tau_D$ determined in this way can be written in the
same general form as used previously[\ref{schacht-optimal_sampling}]
\begin{eqnarray*}
  \sigma_{\tau} & = & f_{\tau} \tau_D \sigma^{( N)}_s = f_{\tau} \tau_D
  \sigma_s \sqrt{\frac{t_{\tmop{trial}}}{T}}
\end{eqnarray*}
The trial time $t_{\tmop{trial}}$ will depend on the wait time used for a
particular trial, but will be of order $\tau_D$. Grouping the exact trial time
with $f_{\tau}$ gives
\begin{eqnarray*}
  \sigma_{\tau} & = & \bar{f}_{\tau} \tau_D \sigma_s \sqrt{\frac{\tau_D}{T}}\\
  \bar{f}_{\tau} & \equiv & f_{\tau} \sqrt{\frac{t_{\tmop{trial}}}{\tau_D}}
\end{eqnarray*}
The previously determined parameters $\sigma_s \approx 0.66$ and $\tau_D
\approx 60 \tmop{ms}$ then gives
\[ \sigma_{\tau} = \bar{f}_{\tau} \frac{0.16 \tmop{ms}}{\sqrt{T /
   \tmop{hour}}} \]
A general optimal sampling analysis[\ref{schacht-optimal_sampling}] gives
$\bar{f}_{\tau}$ as a function of a chosen set of wait times $t_i$ and the
fraction of the total number of trials that each particular wait time is used,
$f_i$. These $t_i$ and $f_i$ can then be chosen to minimize $\bar{f}_{\tau}
\nosymbol$. The optimal results depend implicitly on other details of the
measurement sequence through the total trial time. In practice the trial time
will be given by the wait time plus some fixed overhead $t_0$ that includes
pump and probe times, between trial cooling times, and the other fixed time
blocks described previously that are used for calibration and stability,
giving $t_{\tmop{trial}} = t_0 + t_{\tmop{wait}}$. For these
measurements\tmtextbf{ $t_0 \approx 120 \tmop{ms} \approx 2 \tau_D$}.

For the uniformly sampled case $f_i = 1 / 3$ the optimal result for the
parameters of this system turns out to be
\begin{eqnarray*}
  t^{\tmop{optimal}}_i & = & \{ 0, 0.89, 5.16 \} \tau_D\\
  \bar{f}_{\tau} & \lesssim & 12.3
\end{eqnarray*}
This can be improved slightly by allowing arbitrary $f_i$ and the best optimal
sampling set is
\begin{eqnarray*}
  t^{\tmop{optimal}}_i & = & \{ 0, 0.80, 5.69 \} \tau_D\\
  f & = & \{ 0.28, 0.54, 0.18 \}\\
  \bar{f}_{\tau} & \lesssim & 11.0
\end{eqnarray*}
where $t = 0.72 \tau_D$ is sampled more often than the other cases.

Though only three times are strictly required, such a set would give no extra
information that could be used to check for, or correct for systematic errors,
such as parasitic couplings to the $5 D_{3 / 2} ( F = 2)$ that result in a
non-exponential decay profile. With only three sample times such a deviation
could never be detected. To check for such things a more than minimal set of
wait times is sampled that (almost) covers the overall range of times
indicated by the optimal set. In all the data presented this set is $t_i = \{
0, 5, 10, 20, 50, 100, 200 \}$. In some of the data there are also trials of
$d_1$ which is a block like $d_0$ but beginning with an SPump step. As a
result the trial time becomes $t_{\tmop{trial}} = t_0 + 2 t_{\tmop{wait}}$,
$\bar{f}_{\tau} \rightarrow 19.53$ and gives
\[ \sigma_{\tau} = \frac{4.5 \tmop{ms}}{\sqrt{T / \tmop{hour}}} \]

Such a sensitivity is sufficient for this measurement but could be improved.
First the $d_1$ block can be eliminated, giving again $t_{\tmop{trial}} = t_0
+ t$, and $\bar{f}_{\tau} \rightarrow 17.2$. Similarly a typical measurement
also includes the usual $c_1$, $c_{1 a}$, $n_0$, $n_1$, $c_0$. This provides
useful complementary information for stabilization and systematics, but could
be sampled less often, or omitted if the effects they are used for can be
eliminated or determined by other means. These extra blocks account for most
of the non-$t_{\tmop{wait}}$ trial time. A $d_1$-block-only sequence would
have $t_0 \approx 10 \tmop{ms} = 0.1 \tau_D$ which would further reduce
$\bar{f}_{\tau} \rightarrow 9.8$, improving sensitivity by about a factor of 2
over the currently used measurement scheme.

A more important factor is $n_c$, which directly affects $\sigma_s$.
Increasing $n_c$ by a factor of two decreases $\sigma_s$ by the same factor.
The cooling transition is saturated, so $n_c$ is determined by PMT detection
efficiency and the number of ions. PMT detection efficiency is given mostly by
solid angle and PMT quantum efficiency. Neither is easily improved, but for
future measurements it could be worth the effort if a factor of a few could be
gained. With the present system the number of trapped ions could be increased,
and is for a few cases. But it is not known if multiple ions could affect this
$D$ state lifetime, so most data is taken with a single ion.

\subsection{Lifetime}

\begin{figure}[h]
  a)\resizebox{0.45\columnwidth}{!}{\includegraphics{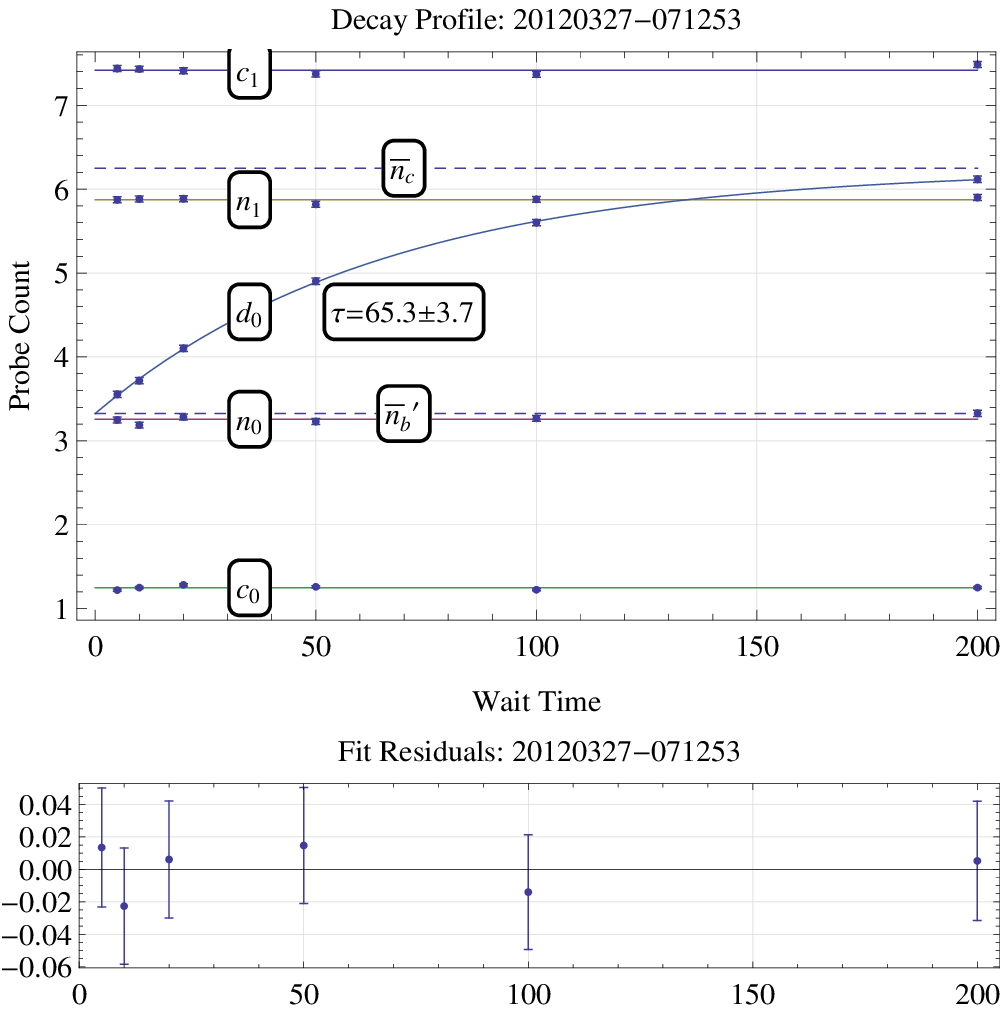}}b)\resizebox{0.45\columnwidth}{!}{\includegraphics{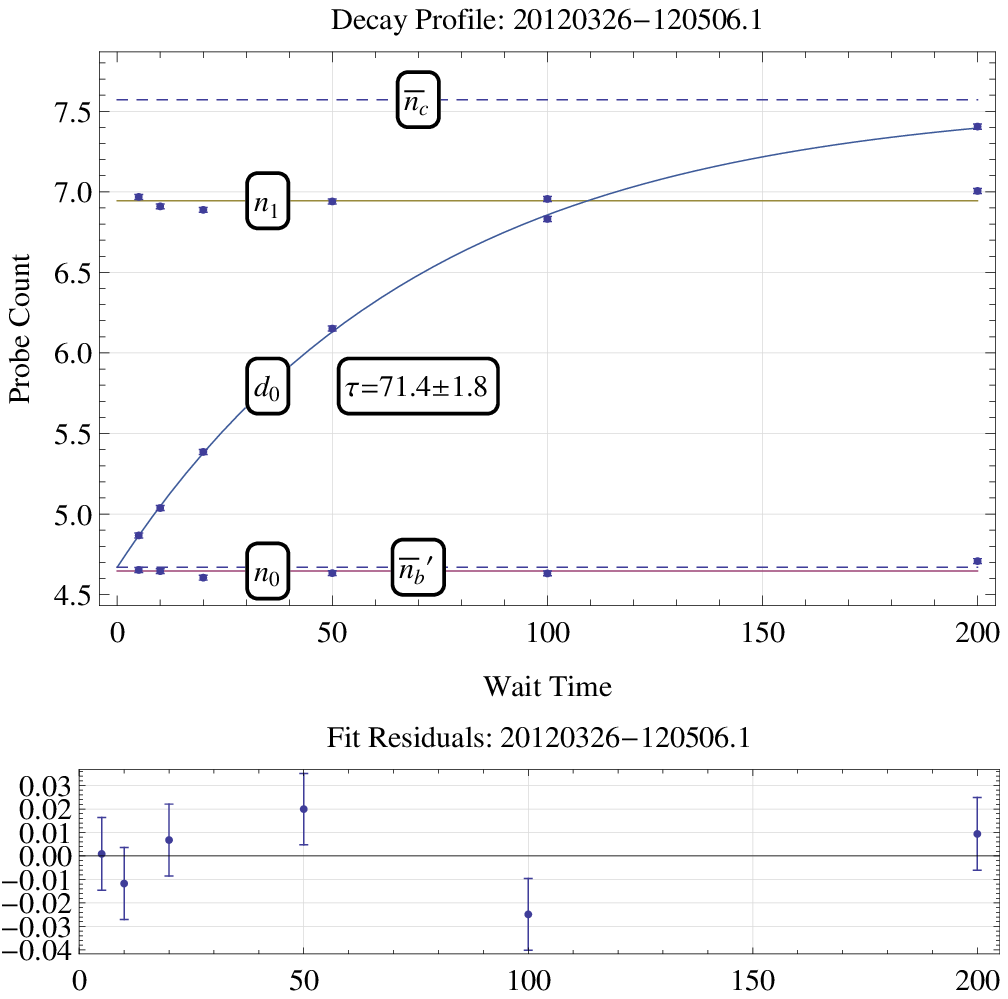}}
  \caption{\label{single-lifetime-profile}Decay profiles from two separate
  data sets with fit residuals. }
\end{figure}

Figure \ref{single-lifetime-profile}a shows $d_1 ( t_{\tmop{wait}})$ for data
already presented in the optimal probe analysis. The $n_i$ and $c_i$ are
independent of $t_{\tmop{wait}}$ indicating a very stable system. The error
derived from the standard deviation of the data collected are smaller than the
plot points used in the profiles, but apparent in the residuals. The profile
fits an exponential decay very precisely. The fit residuals exhibit no
systematic variation from the functional form of the fit model. This
particular data consists of 46690 trials taken over the course of $2.4$ hours.
The uncertainty of $4 \tmop{ms}$ indicates a sensitivity of $5.7 \tmop{ms} /
\sqrt{T / \tmop{hr}}$, similar to the estimate for the sensitivity given
above, indicating that the statistics are generally well understood and that
all the possible sources of variation are identified and accounted for.

Figure \ref{single-lifetime-profile}b shows a second example for the case of
the longest run achieved, and so having the best statistics. This data
consists of $165150$ trials taken over the course of $13.1$ hours. The
uncertainty of $2$ms also gives a similar sensitivity of $6.5 \tmop{ms} /
\sqrt{T / \tmop{hr}}$, which in this case is slightly larger than the $5.0
\tmop{ms} / \sqrt{T / \tmop{hr}}$ that might be expected using the same
considerations as before.

\begin{figure}[h]
  a)\resizebox{0.45\columnwidth}{!}{\includegraphics{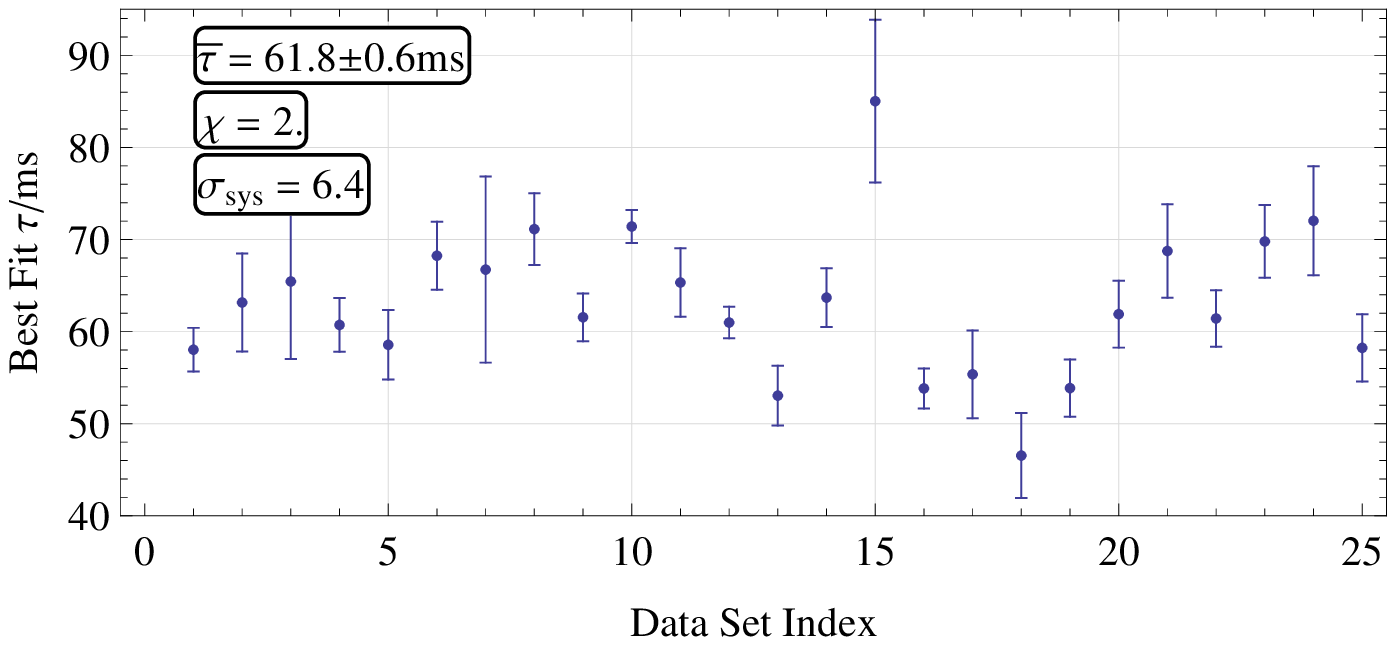}}b)\resizebox{0.45\columnwidth}{!}{\includegraphics{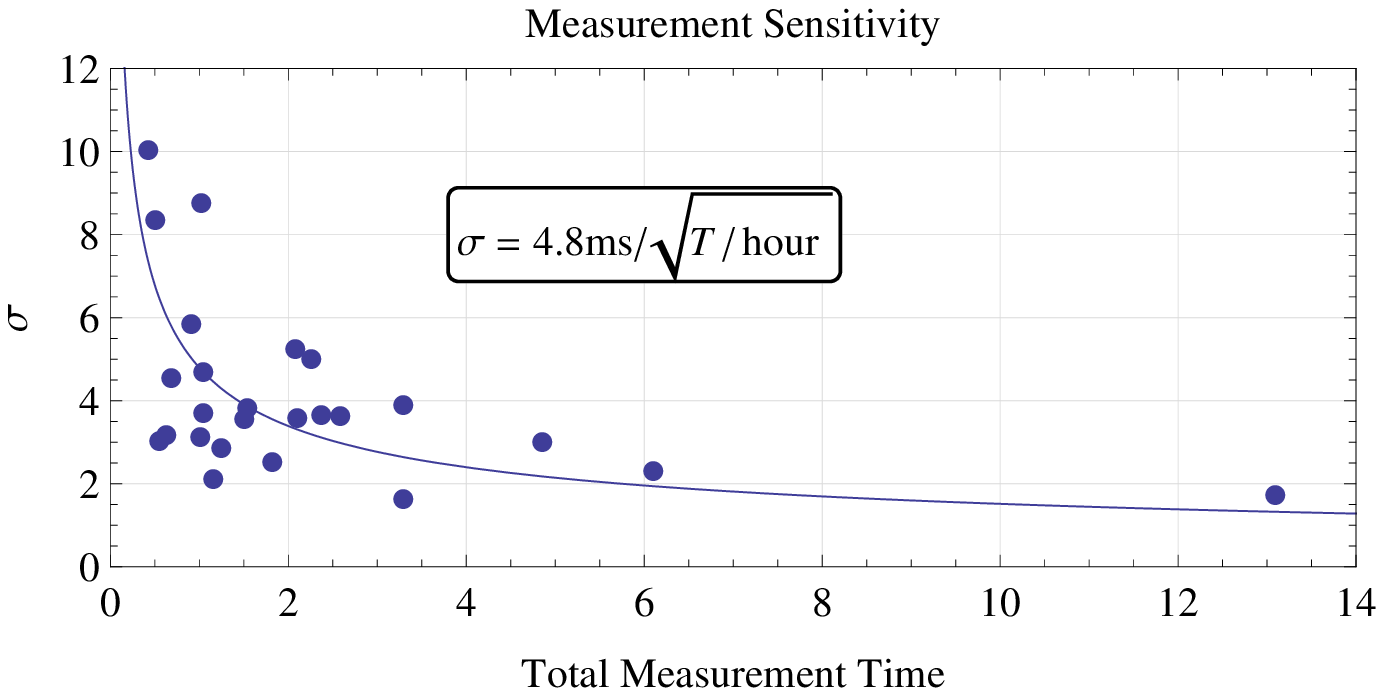}}
  \caption{\label{all-lifetime-data}Lifetime and uncertainty for all lifetime
  data sets with weighted mean, and error as a function of observation time
  and fit to $\sigma \propto 1 / \sqrt{T}$}
\end{figure}

Both of these fits individually are very good, but their results are not
completely consistent with each other, each lying about $1.1 \sigma$ away from
their weighted average. This is not an unreasonable disparity, but larger than
expected, especially given the otherwise very good quality of the statistics,
and more data shows even larger variations. Figure \ref{all-lifetime-data}a
shows the results from all the lifetime data collected for this measurement
including 25 separate runs totaling over $10^6$ trials.

These give a weighted mean of $\bar{\tau} = 61.8 \pm 0.6 \tmop{ms}$. Figure
\ref{all-lifetime-data}b shows the fit $\tau$ uncertainty as a function of the
total measurement time for a particular run. The best fit to $\sigma = a /
\sqrt{T}$ gives a sensitivity of $4.8 \tmop{ms} / \sqrt{T / \tmop{hour}}
\nosymbol$ which is in very good agreement with the statistical analysis.

\subsection{Systematics}

A number of sets exhibit a significant deviation from this mean. An average of
the relative variations from the mean gives
\begin{eqnarray*}
  \chi^2 & = & \frac{1}{N} \sum_i \left( \frac{\tau_i - \bar{\tau}}{\sigma_i}
  \right)^2\\
  \chi & = & 2.0
\end{eqnarray*}
Again, not completely unreasonable but large, and some possible systematic
errors should be considered.

One possibility might be residual couplings from the cooling lasers while
they are nominally off during the wait time that give small extra rates of
excitation to or losses from the $5 D_{3 / 2} ( F = 2)$ state. These rates
have already been considered in the pump and probe analysis and also result in
an exactly exponential decay profile with particular rates. Since these rates
have been seen to vary between measurements, they may then give a shift to the
lifetime that is arbitrary between measurements but stable during a single
measurement. An estimation of their possible size seems to preclude this
possibility. The general sizes of these rates have already been determined to
be on the order of $20 - 30$ Hz when the beams are on. The 370nm laser is
switched by a mechanical shutter that completely blocks the beam when off so
no residual rate extending the lifetime should be expected.

To confirm this, some data was taken with a $d_1 = \tmop{SPump} / \tmop{Off}
( t_{\tmop{wait}}) / \tmop{DProbe}$ block. This block begins with the ion in
the $S$ state. Leaking $370 \tmop{nm}$ light would excite the ion to the $D$
state and the $d_1 ( t_{\tmop{wait}})$ profile should show a decaying time
dependent probe count with the same time constant as the $d_0$ profile. The
asymptotic value of this decay gives the ratio of this possible parasitic
excitation rate to the total rate. This sort of measurement indicates that
this possible spurious excitation rate is statistically consistent with zero
and at worst can be no more than a few 0.1\% of the decay rate.

The 935nm laser is switched by an AOM that is known to be imperfect but still
provides at least $30 \tmop{dB}$ contrast between states. This would allow for
an extra loss rate of $\lesssim 20 - 30 \tmop{mHz}$, shortening the lifetime
an undetectable $0.1 \tmop{ms}$ and not enough to account for the apparent
systematic variation.

Pressure effects should similarly be negligible. Collisional quenching has
been determined to be on the order of $10^7 \tmop{Hz} /
\tmop{torr}$[\ref{yu-lifetime}]. These experiments were all done with
pressures in the low $10^{- 11} \tmop{torr}$ range giving possible reductions
of the lifetime on the order of a few $\tmop{mHz}$ relative to the order
1/50ms=20Hz radiative decay lifetime, which would also be undetectable.

\begin{figure}[h]
  a)\resizebox{0.45\columnwidth}{!}{\includegraphics{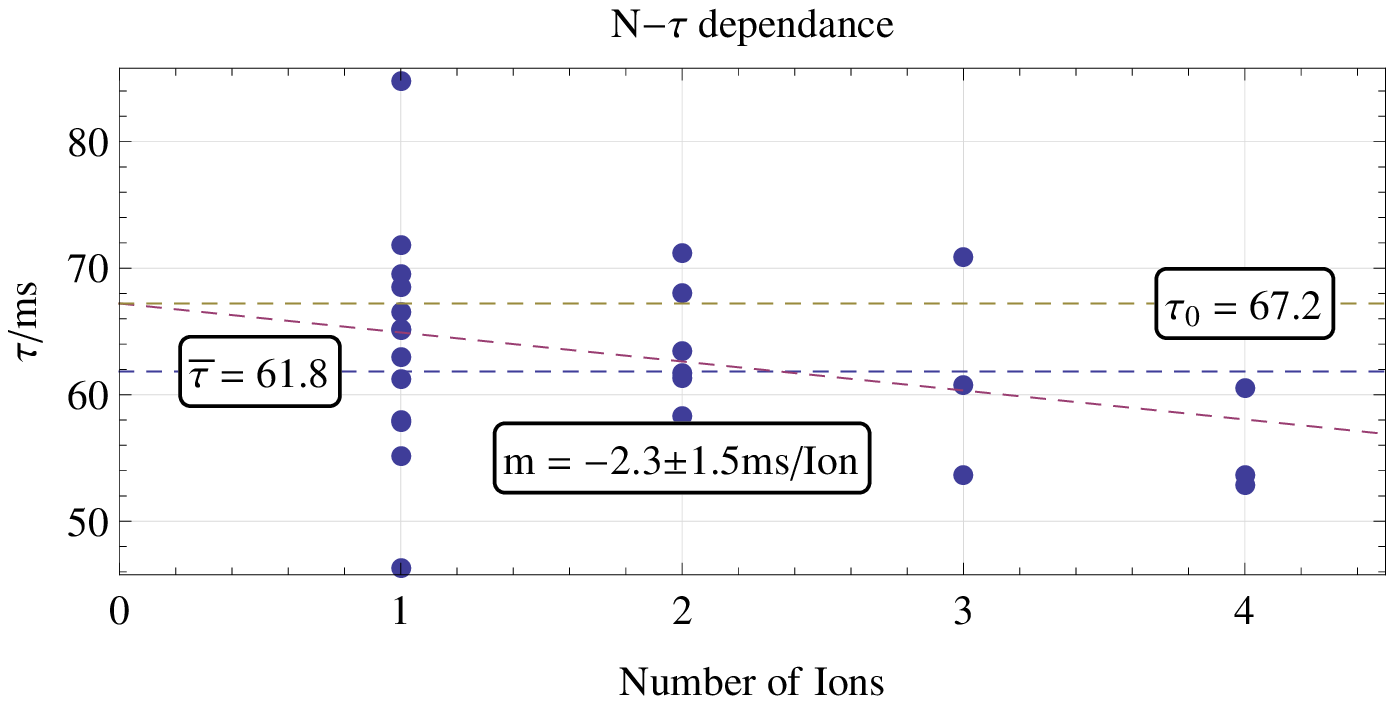}}b)\resizebox{0.45\columnwidth}{!}{\includegraphics{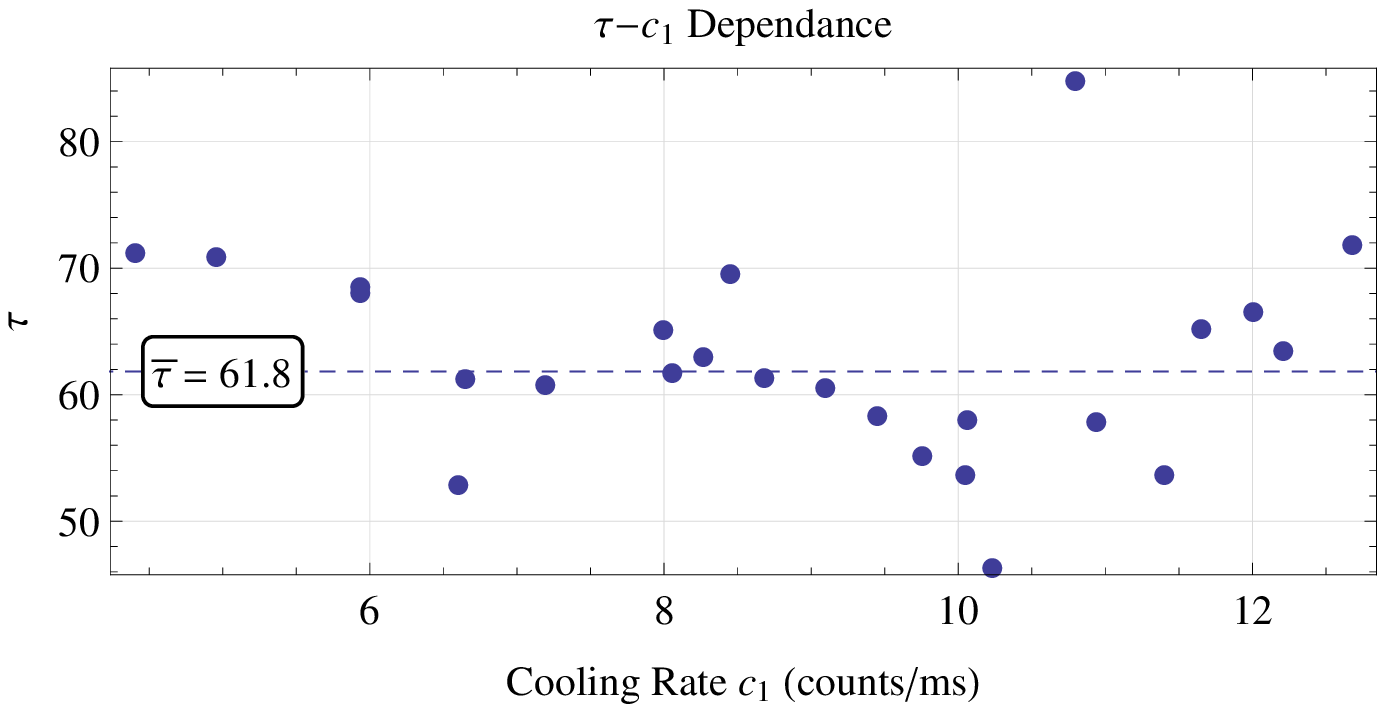}}
  \caption{\label{nc-dependance}Derived lifetime as a function of number of
  ions with linear fit and lifetime as a function of cooling rate.}
\end{figure}

Other possibilities include the number of trapped ions and their temperature.
The mechanism is not clear, but coulomb interactions between the trapped ions,
or details of the micro-motion or secular motion of the ion, or even
super-radiance might individually or in combination provide some coupling to
the ground state that reduces the lifetime. Figure \ref{nc-dependance} shows
the derived lifetime from each data set as a function of the number of ions in
the trap during that run or the cooling rate $c_1$ as a measure of
temperature.

Figure \ref{nc-dependance}a shows a possible dependence on the number of ions
of $- 2.3 \tmop{ms} / \tmop{Ion}$, but it is barely statistically significant,
and the variations within results for the same number of ions are just as
large as the overall variation. $\chi$ improves insignificantly to 1.9. Figure
\ref{nc-dependance}b shows no dependence on cooling rate.

These variations might still be statistical variations and disappear with more
data, but it is just as likely that there is a real systematic variation
between runs. If there is such a shift it appears to be stable during a single
run as sensitivities are completely consistent with statistics, so it would
have to be something that changes when a new ion is loaded. Candidates for
such an effect beyond those few already seemingly ruled out are not apparent.
The variation remains with a size estimated by
\begin{eqnarray*}
  \sigma_{\tmop{sys}}^2 & = & \sum_i \left( \frac{\tau_i -
  \bar{\tau}}{\sigma_i} \right)^2 / \sum_i \frac{1}{\sigma_i^2}\\
  \sigma_{\tmop{sys}} & = & 6.4 \tmop{ms}
\end{eqnarray*}
and gives
\begin{eqnarray*}
  \tau & = & 61.8 \tmop{ms} \pm ( 0.6)_{\tmop{stat}} \pm ( 6.4)_{\tmop{sys}}
\end{eqnarray*}

\section{F State Shelving}

Detection sensitivity is somewhat limited by the shelved state lifetime which
is a modest $\sim 50 \tmop{ms}$, rather shorter than the $100 s +$ seen in
some states in other alkali-like systems. $^{171} \tmop{Yb}^+$ has a $5 F_{7 /
2}$ state with a lifetime in excess of $6$ years (figure \ref{level-diagram}),
which would be an enormous improvement if it could be used as a shelved state.

This $F$ state has an even lower energy than the $5 D$ state, but various
combinations of angular momentum and parity selection rules, and small energy
differences prevent a quick transition to it from either the $5 D$ state or
the $6 P$ state. It is believed that transitions to it have been observed
through collisional couplings to an intermediate state in systems with higher
pressures[\ref{torgerson-yb_f_state}]. In such cases a 638nm laser is used to
clean out the $F$ state through an intermediate $^1 [ 5 / 2]_{5 / 2}$ state.

Driving shelving transitions to this $F$ state can be done in a similar way.
Direct transitions from the ground state would be very difficult, and slow
given they very small coupling indicated by the lifetime. A 410nm laser can
drive a transition from the $5 D$ state to a different intermediate $^1 [ 5 /
2]_{5 / 2}$ state that decays via an E2 transition quickly and principally to
the $F$ state.

A 410nm diode laser was built for this purpose and when applied to the ion
yielded the expected results. While monitoring the cooling signal, the 410nm
laser is applied and the cooling signal is observed to disappear almost
immediately suggesting a successful transition to the $F$ state. The long $F$
state lifetime prevents seeing the radiative decay, but the cooling signal is
immediately restored when the 638nm laser is applied which drives the ion back
into its cooling cycle.

Since this $F$ state is very weakly coupled to any part of the cooling cycle,
the resulting $\tau_{\tmop{probe}}$ should be determined completely by the $F$
state lifetime, $\tau_F$. This gives $a_{\max} \rightarrow 0$ and $\sigma_s$
decreases from $0.66$ to $0.5$its smallest possible value corresponding to
binomial statistics, a modest $25$\% improvement. Such effort for this gain
would probably not be justified for a lifetime measurement which is of limited
interest. But the \ improvement would be welcome in a parity non-conservation
experiment that is statistics-limited where this would correspond directly to
a 25\% improvement in precision. It would also allow for a threshold probe
that is less sensitive to fluctuating experimental parameters. More
importantly $F$ state shelving would also make it practical to do these kinds
of measurements in the isotopes of $\tmop{Yb}$ with zero nuclear spin where
the lack of hyperfine structure prevents the use of the measurement scheme
presented in this article.

\section{Conclusion}

Characterization of the probe coherence times, and careful choices for probe
times give a probe sensitivity close to the maximal possible for an ideal
system even for the fairly poor counting statistics exhibited here. The
specific results shown here for the $D$ state, and the same methods applied to
the $S$ state would allow for significantly improved sensitivity for measuring
the resonance transition frequencies and lightshifts needed for an Atomic
Parity Violation experiment.

For some kinds of measurements, sensitivity to determining a desired quantity
is also affected by pump efficiency and the choice of drive parameters.
Sensitivity can be maximized by characterizing pump times and by using optimal
sampling to determine and optimal set of drive parameters.

Applying these methods to the $D$ state lifetime gives
\begin{eqnarray*}
  \tau & = & 61.8 \tmop{ms} \pm ( 0.6)_{\tmop{stat}} \pm ( 6.4)_{\tmop{sys}}
\end{eqnarray*}
significantly longer than previous measurements and with about 15 times better
statistical precision. The sensitivity this represents of $4.5 \tmop{ms}
\sqrt{\tmop{hour}}$ is also likely very good but this can not be directly
compared to other experiments as total observation time is not commonly
reported.

This kind of precision would be an excellent target to precise atomic
structure calculations, but a larger systematic variation is exhibited that is
very likely to have appeared in previous measurements as well. Though the
source of this apparent systematic variation remains unidentified, the result
for the lifetime is still clearly much longer than previous measurements. As
systematic effects are most likely to reduce the lifetime it is possible that
the these measurements were shorter due to the same effects and didn't have
sufficient sensitivity to resolve similar systematic variations.

In the case of [\ref{yu-lifetime}], which finds $\tau = 52.7 \pm 2.4
\tmop{ms}$ for $^{174} \tmop{Yb}$, the estimated systematic error here is
larger than the $2.4 \tmop{ms}$ quoted uncertainty, but this uncertainty
appears to be underestimated. Using data from Figure 4 of [\ref{yu-lifetime}]
a weighted fit gives a similar $\tau = 52.4 \tmop{ms}$. An estimate of the
variance that neglects the errors of the individual data points gives $\sigma
\approx 2 \tmop{ms}$, also consistent with the published result, but including
those errors in the variance estimate gives $\sigma \approx 9.5 \tmop{ms}$.
This is well outside the range of the estimated systematic variations, and so
this previous measurement may include similar effects without having been able
to resolve them. In that case many trials were made using a single wait time
and experimental parameters could have changed while switching to different
wait times.

If all possible effects are assumed to shorten the lifetime than the best
estimate might more correctly taken to be the largest statistically
significant value found $\tau > \tau_{\max} = 71.4 \pm 1.8 \tmop{ms}$. Which
is approaching the largest calculated result of 74ms[\ref{werth}].

$F$ state shelving would allow for similar state detection lifetime
measurement sensitivity in isotopes with nuclear spin zero, and provide a
direct comparison to previous results as well as testing for possible isotope
dependence.

{\noindent}\tmtextbf{Acknowledgments. }This work was supported by the
Laboratory Directed Research and Development program at Los Alamos National
Laboratory, operated by Los Alamos National Security, LLC for the NNSA U.S.
Department of Energy under contract No.
DE-AC52-06NA25396{\hspace*{\fill}}{\medskip}

\section{\label{h.24nt4umkr5i9}References}

\begin{enumerate}
  \item \label{fortson-ionpnc}N. Fortson
  
  ``Possibility of measuring parity non-conservation with a single trapped
  atomic ion''
  
  19 April 1993, Physical Review Letters 70(16):2383-2386
  
  \item \label{torgerson-ionpnc}J. Torgerson, M. Schacht, J. Zhang
  
  {\textquotedblleft}Measurement of Parity Violation with Single Yb+ Ions''
  
  24 July 2010, Variations of Constants and Violations of Symmetries Workshop,
  Cairns
  
  \item \label{yu-lifetime}N. Yu and L. Maleki
  
  ``Lifetime measurements of the $4 f^{14} 5 d$ metastable states in single
  ytterbium ions''
  
  12 January 2000, Physical Review A 61(2):022507
  
  \item \label{wilson}B.C. Fawcett, M. Wilson
  
  ``Computer Oscillator Strengths, Lande g values, and lifetimes in Yb II''
  
  March 1991
  
  Atomic Data and Nuclear Data Tables 47(2):241-317
  
  \item \label{werth}Ch. Gerz, J. Roths, F. Vedel, G. Werth
  
  ``Lifetime and collisional depopulation of the metastable 5D 3/2-state of
  Yb+''
  
  1988
  
  Zeitschrift f{\"u}r Physik D Atoms, Molecules and Clusters 8(3):235-237
  
  \item \label{dehmelt-shelving}W Nagourney, J Sandberg, H Dehmelt
  
  ``Shelved optical electron amplifier: Observation of quantum jumps''
  
  30 June 1986, Physical Review Letters 56(26):2797-2799.
  
  \item \label{schacht-shelving}M. Schacht, M. Schauer
  
  ``Shelving and Probe Efficiency in Trapped Ion Experiments''
  
  \item \label{schacht-optimal_sampling}M. Schacht,
  
  ``Sensitivity and Optimal Sampling in Precision Experiments''
  
  \item \label{torgerson-yb2+}M M Schauer, J R Danielson, D Feldbaum, M S
  Rahaman, L.-B Wang, J Zhang, X Zhao, J R Torgerson
  
  ``Isotope-selective trapping of doubly charged Yb ions'
  
  27 December 2010, Physical Review A 82(6):062518
  
  \item \label{torgerson-yb_f_state}M M Schauer, J R Danielson, A.-T Nguyen,
  L.-B Wang, X Zhao, J R Torgerson
  
  ``Collisional population transfer in trapped Yb + ions''
  
  5 June 2009, Physical Review A 79(6):062705

\end{enumerate}

\end{document}